\documentclass[prd,aps,twocolumn,reprint,superscriptaddress,nofootinbib,floatfix]{revtex4-2}

\usepackage{mathrsfs}
\usepackage{amsfonts}
\usepackage{amsmath}
\usepackage{mathtools}
\usepackage{xcolor}
\usepackage{units}
\usepackage{amssymb}
\usepackage{graphicx}
\usepackage{subcaption}
\usepackage{bm}
\usepackage{color}
\usepackage{braket}
\usepackage{derivative}
\usepackage{slashed}

\DeclareMathOperator{\tr}{tr}

\newcommand{\vb}[1]{\boldsymbol{#1}}

\begin{document}

\title{Estimates for rare three-body decays of the Omega baryon using chiral symmetry and the $\Delta I = 1/2$ rule}
\author{Cornelis J. G. Mommers}
\affiliation{Institutionen f\"or fysik och astronomi, Uppsala universitet, Box 516, S-75120 Uppsala, Sweden}
\author{Stefan Leupold}
\email{stefan.leupold@physics.uu.se}
\affiliation{Institutionen f\"or fysik och astronomi, Uppsala universitet, Box 516, S-75120 Uppsala, Sweden}
\date{\today}
\begin{abstract}
We study rare three-body decays of the Omega baryon using SU(3) chiral perturbation theory, the successful effective field theory of quantum chromodynamics at low energies. At leading order, we calculate the branching fractions of the decay $\Omega^- \to \Xi \pi \pi$ for all possible combinations of pions. For one channel we find an order-of-magnitude discrepancy between theory and experiment. This tension is known to exist in the non-relativistic limit, and we confirm that it remains in the relativistic calculation. Fairly independent of the values of the low-energy constants we establish lower limits for the branching fractions of these three-body Omega decays, which reaffirm the gap between theory and experiment. We point out that this discrepancy is closely tied to the $\Delta I =1/2$ selection rule. In turn, this means that the three-body decays constitute an interesting tool to scrutinize the selection rule. Using next-to-leading order calculations we also provide predictions for the decay $\Omega^- \to \Xi^0 \mu^- \bar{\nu}_\mu$. We show that fully-differential distributions will provide access to low-energy constants needed in the axial-vector transitions from a decuplet to octet baryon. Since data for all of these rare three-body Omega decays are scarce (fully differential data are nonexistent), we recommend that they be remeasured at running and upcoming experiments, such as BESIII, LHCb, Belle-II, and PANDA. 
\end{abstract}

\maketitle

\section{Introduction and summary}

The $\Omega^-$ baryon acquires a special role in the lowest-lying spin-3/2 decuplet of flavor SU(3). 
While the decays of all other decuplet states are dominated by the strong interaction, the $\Omega^-$ baryon can only decay 
weakly. Weak decays of hadrons are interesting for at least two reasons. First, they can be used as a tool to study CP violation (see Refs.~\cite{bigi-CP,Donoghue:1992dd,Cirigliano:2011ny,Salone:2022lpt} and references therein). 
Here, the overarching context is the search for physics beyond 
the Standard Model. Second, the electroweak interactions constitute an important probe to study the structure of hadrons. Here, 
the overarching context is the study of the poorly-understood non-perturbative sector of the strong interaction. 

Most weakly-decaying baryons are spin-1/2 particles, as photon radiation is usually possible, which de-excites the higher spin state. However, when the minimal quark content is pure in flavor ($qqq$ with $q=u,d,s,c,\ldots$) the Pauli principle together with the color-white structure demand a completely symmetric spin state, i.e.\ spin 3/2 for the ground-state baryon. 
The lowest-lying $uuu$ ($\Delta^{++}$) and $ddd$ ($\Delta^-$) states are heavy enough to allow for the 
strong decay into a nucleon and a Goldstone boson (a pion). But, the $\Omega^-$ baryon, with its three strange quarks, is too light to decay into a cascade ($\Xi$) and an antikaon \cite{pdg}. Pure flavor baryons in the charm or bottom sector are expected to exist, but have not yet been 
found. Thus, the $\Omega^-$ baryon is special. Its decays provide information about the interplay of the strong and 
weak interaction \cite{Goswami:1970wqc,Carone:1991ni,Tandean:1998ch,Egolf:1998vj,Antipin:2007ya} that is complementary 
to the weak decays of the ground state spin-1/2 octet baryons. 

In fact, there are (at least) two puzzles related to the decays of the $\Omega^-$ baryon. 
According to the present data situation, the two-body decays $\Omega^- \to \Xi^- \pi^0, \, \Xi^0 \pi^-$ seem 
to violate the much celebrated isospin selection, or $\Delta I=1/2$ rule \cite{Bourquin19841,pdg}
and the three-body decay $\Omega^- \to \Xi^- \pi^- \pi^+$ seems to show no trace of the $\Xi(1530)$ resonance as an intermediate state in the $\Xi \pi$ channel \cite{HyperCP:2010ego}. 

From a theory point of view, the main purpose of this paper is to show that these two puzzles are interrelated. We should point out right away 
that our purpose is not to solve the two puzzles. 
Instead, we will discuss the consequences of the $\Delta I=1/2$ rule for the three-body decays $\Omega^- \to \Xi^- \pi^+ \pi^-, \, \Xi^- \pi^0 \pi^0$ and $\Xi^0 \pi^- \pi^0$. Partly, the discussion serves to sharpen the contrast between the current experimental situation and the current understanding of the weak decays of hadrons. Based on our findings, we shall argue that differential data on the three-body decays $\Omega^- \to \Xi \pi \pi$ would offer much more information than the two-body decays, if it really turns out that 
the $\Delta I=1/2$ rule is grossly violated in the decays of the $\Omega^-$ baryon. 

Unfortunately, experimental opportunities to study $\Omega^-$ decays, and in extension their potential violation of the $\Delta I = 1/2$ rule, have long been absent. This absence is over now, as $\Omega^-$ baryons (and other hyperons) are and will be abundantly produced in several running and upcoming experiments, 
such as BESIII, LHCb, Belle-II and PANDA \cite{Asner2008,BESIII2020,BESIII:2020lkm,Belyaev2021,Belle-II2010,Belle-II:2018jsg,PANDA2009,PANDA2021}. 
These experiments offer the possibility to study weak baryon decays in significantly more detail. 
Therefore, from an experimental point of view, we regard our paper as a timely reminder of previous discussions, 
where we also deepen the present discourse by interrelating two- and three-body decays. 
Our work is meant as a motivation to 
study, in particular, the differential distributions of three-body decays, which provide access to the 
low-energy constants (LECs) of the effective field theory for the strong and weak interaction in the flavor SU(3) sector. 
Besides leading-order (LO) calculations for $\Omega^- \to \Xi \pi \pi$ in a relativistic framework of 
chiral perturbation theory ($\chi$PT), we will also provide predictions based on next-to-leading order (NLO) 
calculations for the so-far unobserved decay $\Omega^- \to \Xi^0 \mu^- \bar\nu_\mu$. Such semileptonic decays give access to LECs 
that are SU(3)-flavor related \cite{Holmberg:2018dtv,Holmberg:2019ltw} to the production of $\Delta$ baryons 
in neutrino-nucleon and neutrino-nucleus reactions \cite{NuSTEC2017,Procura2008,Geng2008,Mosel2015,Unal2018}.

Let us now briefly review the isospin selection rule, after which we will be in a position to summarize the scope of this paper. In the Standard Model, weak decays of hadrons are driven by an effective four-quark operator of left-handed quarks \cite{Antonelli:1995nv,Cirigliano:2011ny}. For the decays of strange quarks, this transition operator can be decomposed into a left-handed octet and a left-handed 27-plet of flavor SU(3). Assuming octet dominance leads to the $\Delta I = 1/2$ selection rule for nonleptonic weak decays of strange hadrons, like kaons and hyperons. In principle, all possibilities caused by the octet and 27-plet can be encoded in the LECs of a
corresponding hadronic effective field theory (chiral perturbation teory) \cite{Cirigliano:2011ny,Tandean:1998ch}. Yet,
if the number of LECs is large and the data situation poor, the predictive power will be rather limited. In this case,
it is interesting to figure out if some LECs are smaller than others. For a corresponding situation in the strong-interaction
sector of mesons see the discussions in Refs.~\cite{Gasser:1984gg,Ecker:1988te,Donoghue:1988ed}. How the situation can change over the
years with better data and improvements in theory can be deduced from Ref.~\cite{Bijnens:2014lea}.

Coming back to the weak non-leptonic decays of hadrons, one needs better (e.g.\ more differential) data to determine the LECs
or some more microscopic input (lattice QCD, hadronic models, quark models) to reduce the number of LECs by focusing on the
numerically dominant ones.
Another way of describing the microscopic situation is to consider the strange quark that decays. It might either flip directly 
to a down quark (pictured, for instance, by the famous penguin diagram from Ref.~\cite{Ellis:1977uk}) 
or it might produce a down, an up and an 
anti-up quark. In the first case, the isospin changes definitely by $\Delta I = 1/2$. 
In the second case, the three quarks might form an isospin 1/2 or 3/2 combination. In some sense the $\Delta I = 1/2$ selection
rule is based on the simplest possible case. Of course, it is the (poorly understood) hadron structure that dictates which case
of transition happens with which probability amplitude. Therefore, the weak decays offer an opportunity to explore
QCD in the non-perturbative regime.

Phenomenologically, the $\Delta I = 1/2$ rule is in general well satisfied, see Refs.~\cite{Jenkins:1991bt,Tandean:1998ch,Cirigliano:2011ny,Salone:2022lpt} and references therein. For instance, $\Delta I=1/2$ implies that the ratios $\Gamma\left(\Lambda \to p \pi^- \right)/\Gamma\left(\Lambda \to n \pi^0\right)$ and $\Gamma\left( \Xi^- \to \Lambda \pi^-\right)/\Gamma\left(\Xi^0 \to \Lambda \pi^0 \right)$ should both have a value close to $2$, while $\Delta I = 3/2$ implies that the ratios should have a value close to $0.5$. The measured values \cite{pdg}, both approximately $1.77$, clearly favor octet dominance.

However, according to the current experimental status, the selection rule is violated to an unexpectedly large extent by the decays $\Omega^- \to \Xi \pi$. The $\Delta I = 1/2$ rule implies 
\begin{eqnarray}
  \label{eq:Omega-DeltaI12}
  \left. \frac{\Gamma\left(\Omega^- \to \Xi^0 \pi^-\right)}{\Gamma\left(\Omega^- \to \Xi^- \pi^0 \right)} \right\vert_{\rm theory} 
  \approx 2
\end{eqnarray}
(while pure $\Delta I = 3/2$ implies that the ratio would be approximately equal to $0.5$). 
Measurements give approximately \cite{Bourquin19841,pdg}
\begin{eqnarray}
  \label{eq:Omega-DeltaI12-exp}
  \left. \frac{\Gamma\left(\Omega^- \to \Xi^0 \pi^-\right)}{\Gamma\left(\Omega^- \to \Xi^- \pi^0 \right)} \right\vert_{\rm exp.} 
  \approx 2.74  \,,
\end{eqnarray}
which suggests an unexpectedly strong admixture of a 
$\Delta I = 3/2$ amplitude, interfering positively for $\Xi^0 \pi^-$ and negatively for $\Xi^- \pi^0$. If this finding is 
confirmed by new measurements, the current effective-field-theory picture about the interplay of the strong and the weak interaction will receive significant modifications. 

On the theoretical side we can follow two lines of thought. On the one hand, we can explore the consequences of abandoning the $\Delta I = 1/2$ rule. This has been done in Ref.~\cite{Tandean:1998ch}, with the conclusion that more data are needed to make phenomenologically useful predictions. On the other hand, we can posit the $\Delta I = 1/2$ rule to be true and look for more disagreements between theory and experiment. This work follows the second line of reasoning. In particular, we look at three-body decays, as differential data give much more information than just comparing two numbers (2 versus 2.74 for the previously discussed ratio). 

We will present new calculations for the decays $\Omega^- \to \Xi \pi \pi$ (for all combinations of pions). These calculations will be kept at LO, as there are not enough data to pin down the LECs that appear at NLO. Moreover, given the scanty data, an NLO calculation is of little added value. In the context of heavy-baryon $\chi$PT, the decay width for $\Omega^- \to \Xi^- \pi^+ \pi^-$ has already been calculated in Ref.~\cite{Antipin:2007ya}. We will extend this calculation by including more decays, calculating fully differential distributions and supplanting the heavy-baryon approximation with a fully relativistic formalism. We will put special emphasis on the $\Delta I = 1/2$ rule and use it, together with chiral symmetry, to provide model-independent lower bounds for the branching fractions. What makes our lower bounds robust is that chiral symmetry breaking relates the decay channels $\Xi \pi \pi$ and $\Xi \pi$. We will find an order-of-magnitude disagreement between current experimental data and our robust lower bound for the measured channel $\Omega^- \to \Xi^- \pi^+ \pi^-$. Even though all building blocks to establish this lower bound have existed in the literature, we are not aware that it has been spelled out that the failure to understand the current data of $\Omega^- \to \Xi^- \pi^+ \pi^-$ provides a further challenge to the $\Delta I = 1/2$ rule. In turn, this means that such three-body decays $\Omega^- \to \Xi \pi \pi$ could be used to scrutinize our understanding of the strong and weak interaction. The general theme of this work is that three-body decays $\Omega^- \to \Xi \pi \pi$, $\Omega^- \to \Xi \mu \bar \nu_\mu$, and 
$\Omega^- \to \Xi e \bar \nu_e$ (see also Ref.~\cite{Holmberg:2019ltw}) offer the opportunity to determine LECs of the effective field 
theory of QCD at the low energies probed by hyperon decays. 

The rest of this work is structured in the following way: in the next section we specify the relativistic chiral Lagrangian 
that describes at LO the strong and weak interactions of the baryon decuplet, the baryon octet and the Goldstone boson octet. 
We supplement this Lagrangian by the subleading NLO terms that we need in the present work. We also specify our power 
counting \cite{Holmberg:2018dtv,Holmberg:2019ltw}, i.e.\ we give a meaning to the phrase ``NLO''. 
In Sec.\ \ref{sec:results} we utilize the LO Lagrangian to present the decays $\Omega^- \to \Xi \pi \pi$ in great detail. 
In Sec.\ \ref{sec:munu}, we provide NLO predictions for the semi-leptonic decay $\Omega^- \to \Xi^0 \mu^- \bar \nu_\mu$. A brief outlook is presented in Sec.\ \ref{sec:outlook}. 
Some technical aspects that would interrupt the main line of reasoning have been relegated to appendices. 

\section{The SU(3) Chiral Lagrangian}
\label{sec:lagr}

The dominant parts of the interactions of octet and decuplet baryons with Goldstone bosons and external fields at low energies 
are provided by the following LO chiral SU(3) Lagrangian \cite{Holmberg:2018dtv,Geng:2008mf,Lutz:2001yb,Semke:2005sn,Pascalutsa:2006up}:
\begin{widetext}
    \begin{eqnarray}
        \mathcal{L}_{\text{baryon}}^{\left(1\right)} &=& \tr\left[ \bar{B} \left( i\slashed{D} - m_{\left(8\right)} \right) B \right] + \bar{T}^\mu_{abc} \left[ i \gamma_{\mu\nu\alpha} \left(D^\alpha T^\nu \right)^{abc} - \gamma_{\mu\nu} m_{\left(10\right)} \left( T^\nu \right)^{abc} \right] - \frac{H_A}{2} \bar{T}^\mu_{abc} \gamma_\nu \gamma_5 \left(u^\nu\right)^c_d T^{abd}_\mu \nonumber \\*
        &&{} + \frac{D}{2} \tr\left( \bar{B} \gamma^\mu  \gamma_5 \left\{ u_\mu, B \right\} \right) + \frac{F}{2} \tr\left( \bar{B} \gamma^\mu \gamma_5 \left[ u_\mu, B \right] \right) + \frac{h_A}{2\sqrt{2}} \left( \epsilon^{ade} \bar{T}^\mu_{abc} \left(u_\mu\right)^b_d B^c_e + \epsilon_{ade} \bar{B}^e_c \left( u^\mu \right)^d_b T^{abc}_\mu \right) \,, 
\label{eq:strong}
    \end{eqnarray}
\end{widetext}
where tr denotes a flavor trace. In passing, we have introduced totally antisymmetric products of gamma matrices,
\begin{eqnarray}
    \gamma^{\mu\nu} &=& \frac{1}{2}\left[ \gamma^\mu, \gamma^\nu \right] = -i\sigma^{\mu\nu}, \\*
    \gamma^{\mu\nu\alpha} &=& \frac{1}{2} \left\{ \gamma^{\mu\nu}, \gamma^\alpha \right\} = i\epsilon^{\mu\nu\alpha\beta} \gamma_\beta \gamma_5. 
\end{eqnarray}
In agreement with Ref.~\cite{pesschr} our conventions are $\gamma_5 = i \gamma^0 \gamma^1 \gamma^2 \gamma^3$ and $\epsilon_{0123} = -1$. 

This Lagrangian is supplemented by the corresponding LO chiral Lagrangian for the Goldstone bosons \cite{Weinberg:1978kz,Gasser:1983yg,Gasser:1984gg,Scherer:2002tk,Scherer:2012xha}:
\begin{eqnarray}
    \mathcal{L}^{\left(2\right)}_{\text{meson}} = \frac{F_\pi^2}{4}\tr\left(u_\mu u^\mu + \chi_+\right) \,.
\label{eq:strong-meson}
\end{eqnarray}
Finally, we add the terms that describe the non-leptonic weak decays at the respective LO for baryons and 
Goldstone bosons \cite{Jenkins:1991bt,Cirigliano:2011ny}: 
\begin{eqnarray}
  \mathcal{L}_{\text{weak}}^{\left(0\right)} &=& 
  h_C \bar{T}^\mu_{abc} \left[ u \left( h + h^\dagger \right) u^\dagger \right]^c_d T_\mu^{abd} \nonumber \\*
  && {} + h_D \tr\left( \bar{B} \left\{ u \left( h + h^\dagger \right) u^\dagger, B \right\} \right) \nonumber \\*
  && {} + h_F \tr\left( \bar{B} \left[ u \left( h + h^\dagger \right) u^\dagger, B \right] \right)   \nonumber \\*
  && {} + \frac{1}{4} h_\pi F_\pi^2 \tr\left[ u \left( h + h^\dagger \right) u^\dagger u_\mu u^\mu\right]  \,.
  \label{eq:weak-mb}
\end{eqnarray}

In the above Lagrangians we have packaged the mesons into
\begin{equation}
  \Phi = \begin{pmatrix} \pi^0 + \frac{1}{\sqrt{3}} \eta & \sqrt{2} \pi^+ & \sqrt{2} K^+ \\ \sqrt{2} \pi^- & -\pi^0 + \frac{1}{\sqrt{3}} \eta & \sqrt{2} K^0 \\ \sqrt{2} K^- & \sqrt{2} \bar{K}^0 & - \frac{2}{\sqrt{3}} \eta  \end{pmatrix}.
  \label{eq:1}
\end{equation}
In our convention $\Phi^a_b$ is the entry in the $a$th row and $b$th column. 
Additionally \cite{Scherer:2002tk,Scherer:2012xha},\footnote{Note that in this parametrization there is a sign difference to \cite{Ecker:1988te}.} 
\begin{equation}
  u^2 = U = \exp\left(i\Phi/F_\pi\right).
  \label{eq:2}
\end{equation}
These Goldstone boson fields can be combined into a source-like term $u_\mu = i u^\dagger \left( \nabla_\mu U \right) u^\dagger = u^\dagger_\mu$.

The spin-1/2 octet baryons are encoded in 
\begin{equation}
    B = \begin{pmatrix}
        \frac{1}{\sqrt{2}} \Sigma^0 + \frac{1}{\sqrt{6}} \Lambda & \Sigma^+ & p \\ \Sigma^- & -\frac{1}{\sqrt{2}} \Sigma^0 + \frac{1}{\sqrt{6}} \Lambda & n\\ \Xi^- & \Xi^0 & -\frac{2}{\sqrt{6}} \Lambda
    \end{pmatrix}.
    \label{eq:3}
\end{equation}
Spin-3/2 decuplet baryons are collected in a totally flavor-symmetric tensor $T^{abc}$, where
\begin{eqnarray}
  &T^{111} = \Delta^{++}, \quad &T^{112} = \frac{1}{\sqrt{3}} \Delta^+, \nonumber \\* 
  &T^{122} = \frac{1}{\sqrt{3}} \Delta^0, \quad &T^{222} = \Delta^-, \nonumber \\*
  &T^{113} = \frac{1}{\sqrt{3}} \Sigma^{\ast +}, \quad &T^{123} = \frac{1}{\sqrt{6}} \Sigma^{\ast 0}, \quad T^{223} = \frac{1}{\sqrt{3}} \Sigma^{\ast -},  \nonumber \\*
  &T^{133} = \frac{1}{\sqrt{3}} \Xi^{\ast 0}, \quad &T^{233} = \frac{1}{\sqrt{3}} \Xi^{\ast -}, \quad  T^{333} = \Omega^-.  
  \label{eq:decuplet-states}
\end{eqnarray}

The chirally-covariant derivative acting on these fields is defined as
\begin{eqnarray}
    D^\mu B &=& \partial^\mu B + \left[\Gamma^\mu, B\right], \\
    \left(D^\mu T\right)^{abc} &=& \partial^\mu T^{abc} + \left(\Gamma^\mu\right)^a_{d} T^{dbc} + \left(\Gamma^\mu\right)^b_{d} T^{adc} \nonumber \\*
    &&\quad + \left(\Gamma^\mu\right)^c_{d} T^{abd},\\
    \left(D^\mu \bar{T}\right)_{abc} &=& \partial^\mu \bar{T}_{abc} -  \left(\Gamma^\mu\right)^{d}_a \bar{T}_{dbc} - \left(\Gamma^\mu\right)^{d}_b \bar{T}_{adc} \nonumber \\*
    &&\quad - \left(\Gamma^\mu \right)^{d}_c \bar{T}_{abd},\\
    \nabla_\mu U &=& \partial_\mu U - i\left(v_\mu + a_\mu\right) U \nonumber \\*
    &&\quad + iU\left(v_\mu - a_\mu\right), 
\end{eqnarray}
where
\begin{eqnarray}
    \Gamma_\mu &=& \frac{1}{2}\big\{ u^\dagger \left[ \partial_\mu - i\left(v_\mu + a_\mu \right) \right] u \nonumber \\*
    &&\quad + u \left[ \partial_\mu - i\left(v_\mu - a_\mu\right) \right] u^\dagger\big\}.
\end{eqnarray}

The external sources can be used to mediate the interactions with the fields that do not couple to the strong interaction \cite{Gasser:1983yg}. In our case these are the weak source $h$ emerging effectively from the left-handed four-quark operator, the vector source $v_\mu$, the axial-vector source $a_\mu$, the scalar source $s$ and the pseudoscalar source $p$. We can combine the latter two into $\chi = 2 B_0 \left( s + ip\right)$, from which we define $\chi_\pm = u^\dagger \chi u^\dagger \pm u \chi^\dagger u$. 

For the nonleptonic weak decays one replaces the elements of the octet matrix $h$ by $h_{ab} = \delta_{a2}\delta_{3b}$ or 
equivalently $h + h^\dagger\mapsto \lambda_6$ \cite{Cirigliano:2011ny}. 
This choice for $h$ selects the $s \to d$ transitions \cite{Jenkins:1991bt} consistent with the $\Delta I = 1/2$ rule. 
We can introduce interactions with electromagnetism via the replacement \cite{Scherer:2002tk,Scherer:2012xha}
\begin{equation}
    v_\mu \mapsto e A_\mu \begin{pmatrix} \frac{2}{3} & 0 & 0 \\ 0 & -\frac{1}{3} & 0 \\ 0 & 0 & -\frac{1}{3}   \end{pmatrix},
    \label{eq:11}
\end{equation}
where $A_\mu$ is the photon field and $e$ the proton charge. 
For semi-leptonic weak decays mediated by $W$-bosons we replace \cite{Scherer:2002tk,Scherer:2012xha}
\begin{equation}
    v_\mu - a_\mu \mapsto -\frac{g_w}{\sqrt{2}} W_\mu^+ \begin{pmatrix} 0 & V_{ud} & V_{us} \\ 0 & 0 & 0 \\ 0 & 0 & 0 \end{pmatrix} + \text{H.c.},
\label{eq:12}
\end{equation}
with the $W$-boson field $W_\mu^+$, the Cabibbo-Kobayashi-Maskawa matrix elements $V_{ud}$ and $V_{us}$ \cite{Kobayashi:1973fv}, and the weak gauge coupling $g_w$.

The pion-decay constant is $F_\pi = 92.4$ MeV, and $H_A$, $D$, $F$, $h_A$, $B_0$, $h_C$, $h_D$, $h_F$, and $h_\pi$ are other LECs. 
Those needed for our purposes will be specified below in Sec.\ \ref{sec:results}. 

The octet and decuplet baryon masses in the chiral limit are $m_{\left(8\right)}$ and $m_{\left(10\right)}$, respectively. 
In practice, we take the physical masses for all the states. Formally, this can be
achieved by adding NLO mass-splitting terms \cite{Holmberg:2018dtv} to the strong Lagrangian \eqref{eq:strong}. 
We will not provide those terms explicitly, but we shall specify the elements of the 
NLO Lagrangian \cite{Holmberg:2019ltw} that are needed for the semi-leptonic decays: 
\begin{eqnarray}
    \mathcal{L}_{\text{baryon}}^{\left(2\right)} &=&  
    i c_M \epsilon_{ade} \bar{B}^e_c \gamma_\mu \gamma_5 \left( f_+^{\mu\nu} \right)^d_b T^{abc}_\nu \nonumber \\*
    &&{} + i c_E \epsilon_{ade} \bar{B}^e_c \gamma_\mu  \left(f_-^{\mu\nu}\right)^d_b T^{abc}_\nu + \text{H.c.} 
\label{eq:strong-NLO-Lagr}
\end{eqnarray}
The new LECs are $c_M$ and $c_E$. The field strengths $f_\pm^{\mu\nu}$ \cite{Ecker:1988te,Bijnens:2014lea} are given by
\begin{equation}
    f_\pm^{\mu\nu} = u F_L^{\mu\nu} u^\dagger \pm u^\dagger F_R^{\mu\nu} u,
    \label{eq:13}
\end{equation}
with
\begin{equation}
    F^{\mu\nu}_{R,L} = \partial^\mu\left(v^\nu \pm a^\nu \right) - \partial^\nu\left( v^\mu \pm a^\mu \right) - i\left[v^\mu \pm a^\mu, v^\nu \pm a^\nu \right].
    \label{eq:14}
\end{equation}

With respect to chiral transformations \cite{Scherer:2012xha,Jenkins:1991es} $L \in \text{SU(3)}_L$, $R \in \text{SU(3)}_R$ and $h_V \in \text{SU(3)}_V$, the fields transform as\footnote{We remark that the transformation properties for $U$ and $u$ given in Refs.~\cite{Holmberg:2018dtv,Holmberg:2019ltw} are incorrect. The authors mixed conventions of Ref.~\cite{Jenkins:1991es} and Ref.~\cite{Scherer:2012xha}; see also Appendix \ref{app:convention}.}
\begin{eqnarray}
    U & \mapsto & RUL^\dagger, \quad u \mapsto Ruh_V^\dagger = h_VuL^\dagger, \nonumber \\*
    u_\mu & \mapsto & h_V u_\mu h_V^\dagger, \quad B \mapsto h_V B h_V^\dagger, \nonumber \\*
    h & \mapsto & L h L^\dagger, \quad u h u^\dagger \mapsto h_V \left( u h u^\dagger \right) h_V^\dagger, \nonumber \\*
    T^{abc}_\mu & \mapsto & \left(h_V\right)^a_d \left(h_V\right)^b_e \left(h_V\right)^c_f T^{def}_\mu, \nonumber \\*
    \bar{T}^\mu_{abc} & \mapsto & \left( h_V^\dagger \right)^d_a \left(h_V^\dagger\right)^e_b \left(h_V^\dagger\right)^f_c \bar{T}^\mu_{def}.
\label{eq:trafo-right}
\end{eqnarray}

Finally, let us further specify our formalism and our power counting. Our Lagrangians are fully relativistic and we shall carry 
out all calculations in a relativistic manner. Conceptually, this only makes sense if it is clear how loop corrections are 
handled in the relativistic framework. 
This requires us to know how important each diagram is in our power-counting scheme. And, it requires that 
all diagrams contributing to a certain order in the power counting can be calculated. Otherwise, predictive power would be lost.
Historically, it was not clear how to achieve these requirements. As a consequence the first systematic scheme that 
was established was the heavy-baryon formalism. Previously, decays of $\Omega^-$ baryons 
had been carried out in this non-relativistic framework \cite{Jenkins:1991bt,Egolf:1998vj,Tandean:1998ch,Antipin:2007ya}. 
Meanwhile, though, it has been established how to perform relativistic calculations that follow a consistent power counting scheme.
Therefore it makes sense that we perform our calculations in a fully relativistic way. 
For a review of all these developments from heavy-baryon to relativistic chiral formulations, we refer to Ref.~\cite{Scherer:2012xha}. We will come back to the power counting in Sec.\ \ref{sec:munu}. 

\section{Non-leptonic decays to cascades and pions}
\label{sec:results}

\subsection{The two-body decays $\Omega^- \to \Xi \pi$ and the consequences of chiral symmetry}
\label{sec:Xipi}

The simplest microscopic process compatible with the $\Delta I = 1/2$ rule is the decay of a strange to a down quark. 
One starts with an $\Omega^-$ state and its minimal quark content $sss$. After the decay one obtains a state with quark content $dss$ and spin 3/2. 
The strong interaction can add extra quark-antiquark pairs. The corresponding multi-hadron states have 
overlap with the $dss$ configuration, but there is also a low-lying single-hadron state (three-quark state) 
with quark content $dss$ and spin 3/2: the 
$\Xi(1530)$ resonance \cite{pdg}, denoted by $\Xi^*$ in Eq.~\eqref{eq:decuplet-states}. Sharing the very same flavor multiplet, 
it sits close by in mass to the decaying  $\Omega^-$ state. Therefore, the $\Xi(1530)$ will contribute significantly to the 
decay process $\Omega^- \to \Xi \pi$ where the final state emerges as a strong decay process from a virtual $\Xi(1530)$.  

This picture of a hadron fluctuating first into another hadron, 
which is also applied to the weak decays of spin-1/2 hyperons \cite{Jenkins:1991bt} and kaons \cite{Cirigliano:2011ny}, 
lies at the heart of the Lagrangian \eqref{eq:weak-mb}. For the case at hand this concerns the term $\sim h_C$. 
\begin{figure}[ht] 
    \centering
        \includegraphics[width=0.65\linewidth]{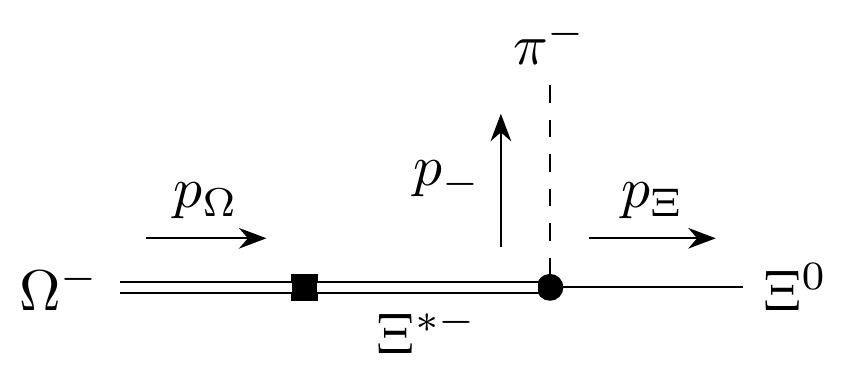}
    \caption{Leading-order diagram for the decay $\Omega \to \Xi \pi$. Circles indicate a strong vertex and squares indicate a weak vertex.}
    \label{fig:feyn-2body}
\end{figure}
These considerations lead to the LO diagram shown in Fig.\ \ref{fig:feyn-2body} 
and to the LO 
formula (see also \cite{Jenkins:1991bt} for the corresponding heavy-baryon calculation)
\begin{eqnarray}
  \label{eq:Om-Xipi-both}
  \lefteqn{\Gamma\left(\Omega^- \to \Xi^0 \pi^-\right) =   2 \Gamma\left(\Omega^- \to \Xi^- \pi^0 \right)} \nonumber \\
  & = & \frac{p_{\text{cm}}}{8 \pi m_\Omega^2} \frac{h_A^2 h_C^2}{432 F_\pi^2} \frac{1}{m_\Omega^2 \left(m_\Omega - m_{\Xi^\ast}\right)^2} \nonumber \\
  & \times & \left(m_\Omega^2 - 2m_\Omega m_\Xi - m_\pi^2 + m_\Xi^2\right) \nonumber \\
  & \times & \left( m_\Omega^2 + 2 m_\Omega m_\Xi - m_\pi^2 + m_\Xi^2\right)^2 \,, \phantom{m}  
\end{eqnarray}
where
\begin{eqnarray}
    p_{\text{cm}} & = & \frac{1}{2m_\Omega} \sqrt{ \left( m_\Omega + m_\pi \right)^2 - m_\Xi^2 } \nonumber \\
    & \times & \sqrt{ \left( m_\Omega - m_\pi \right)^2 - m_\Xi^2 }.
\end{eqnarray}
Here one sees how the $\Xi^* = \Xi(1530)$ resonance enhances the process by the factor 
$1/(m_\Omega-m_{\Xi^*}) \sim 1/M_K^2$ in the amplitude. 

In passing, we note that the process depicted in Fig.~\ref{fig:feyn-2body} describes 
the parity-conserving p-wave amplitude. The parity-violating d-wave is phase-space suppressed \cite{Jenkins:1991bt,Egolf:1998vj}.
More generally, it should be stressed that Eq.~\eqref{eq:Om-Xipi-both} is a result valid at LO. 
There are chiral 
corrections that are, however, hard to control beyond leading-log accuracy \cite{Jenkins:1991bt,Egolf:1998vj} 
due to a plethora of LECs in the weak NLO Lagrangian that 
complements the LO structures of Eq.~\eqref{eq:weak-mb}. Nevertheless, the dominant contribution is given by Eq.~\eqref{eq:Om-Xipi-both}. 

Of course, all this is compatible with the $\Delta I = 1/2$ rule, which amounts to the prediction 
$\Gamma\left(\Omega^- \to \Xi^0 \pi^-\right)/\Gamma\left(\Omega^- \to \Xi^- \pi^0 \right) = 2$.  
As already stressed this is in disagreement with the current data. The results have been used to 
question the data \cite{Jenkins:1991bt,Egolf:1998vj} or to speculate 
on interesting new aspects related to $\Delta I = 3/2$ processes \cite{Carone:1991ni}. In view of this, our perspective is that additional experimental verification of the present results is necessary. In addition, 
one learns even more from three-body decays to which we turn next.
 
As happens very often in low-energy QCD, the interesting aspect is chiral symmetry breaking. 
It relates in a model-independent way the transition amplitudes $A \to B$ to processes with $n$ additional soft pions, $A \to B + n  \pi_\text{soft}$ \cite{Weinberg:1966kf}. In other words, given the amplitude $A \to B$, we can predict {\em without any new parameters} the strength for the amplitude $A \to B + n \pi$ up to chiral corrections. For strong-interaction processes the number of pions must be even to conserve parity. But, for weak processes one predicts a parity-breaking amplitude $A \to B \pi$ from a parity-conserving amplitude $A \to B$, and a parity-conserving amplitude from a parity-breaking one. Diagramatically the processes 
of Fig.~\ref{fig:feyn-2body} and Fig.~\ref{fig:feynd}, diagram 3 are intimately related.

Taking the previous considerations together, the $\Delta I = 1/2$ selection rule and chiral symmetry relate the strength of the parity-conserving part of the amplitude of $\Omega^- \to \Xi \pi$ to the parity-violating part of the amplitude of $\Omega^- \to \Xi \pi \pi$. We will use this connection to provide an estimate for the lower limit of the ratio of decay widths for the processes $\Omega^-\to \Xi \pi \pi$ and $\Omega^- \to \Xi \pi$. 

Technically, the additional soft pion emerges from the appearance of $u=\exp(i\Phi/(2 F_\pi))$ in Eq.~\eqref{eq:weak-mb}. We are 
by no means the first who use this Lagrangian. Yet, to the best of our knowledge it has never been pointed out before that 
the $\Delta I = 1/2$ selection rule can be used for an otherwise model independent prediction of the ratio
$\Gamma(\Omega \to \Xi \pi \pi)/\Gamma(\Omega \to \Xi \pi)$. Instead, it has been stressed in the literature how uncertain 
the determination of $h_C$ and other parameters is. But one can easily overestimate the uncertainties if one does not resort 
to some constraints provided by chiral symmetry. 

Though chiral perturbation theory might not always converge well \cite{Jenkins:1991bt}, 
the order of magnitude of our lower-limit prediction should be regarded as correct. Therefore, we interpret a gross violation of this prediction not merely as a challenge for chiral perturbation theory but rather as a challenge of the $\Delta I = 1/2$ selection rule.

\subsection{The decay $\Omega^- \to \Xi^- \pi^+ \pi^-$}

Meanwhile, the concepts of current algebra and partially conserved axial-vector current \cite{Weinberg:1966kf} are encompassed and 
systematized by the use of the chiral Lagrangians 
\cite{Weinberg:1978kz,Gasser:1983yg,Gasser:1984gg,Jenkins:1991es,Jenkins:1991bt,Scherer:2002tk,Cirigliano:2011ny,Scherer:2012xha,Bijnens:2014lea}. 
We will determine some corrections to the picture that we have just described. However, the small size of these corrections will support our claims.  

We start with the decay $\Omega^- \to \Xi^- \pi^+ \pi^-$. We work with kinematic variables $m^2\left(\Xi^- \pi^\pm\right) = \left(p_\Xi + p_\pm\right)^2$. Alternatively, we may choose $\cos\theta$ as a second variable, instead of $m^2\left(\Xi^- \pi^-\right)$. Here, $\theta$ is the angle between $\vb{p}_\Xi$ and $\vb{p}_-$ in the frame where $\vb{p}_\Xi + \vb{p}_+  = \vb{0}$. The relationship between $\cos\theta$ and $m^2 \left( \Xi^- \pi^- \right)$ is 
    \begin{equation}
        m^2 \left( \Xi^- \pi^- \right) = \left(E^\ast_{\Xi} + E^\ast_-\right)^2 - \vb{p}_\Xi^2 - \vb{p}_-^2 - 2 \lvert \vb{p}_\Xi \rvert \lvert \vb{p}_- \rvert \cos\theta,
    \end{equation}
where
    \begin{eqnarray}
        E^\ast_{\Xi} & = & \frac{m^2\left(\Xi^- \pi^+\right) - m_\pi^2 + m_\Xi^2}{2 m\left(\Xi^- \pi^+\right)} ,\nonumber  \\
        E^\ast_{-} & = & \frac{-m^2\left(\Xi^- \pi^+\right) + m_\Omega^2 - m_\pi^2}{2 m\left(\Xi^- \pi^+\right)} ,\nonumber  \\
        \lvert \vb{p}_\Xi \rvert & = & \frac{\lambda^{1/2}\left(m^2\left(\Xi^- \pi^+\right),m_\Xi^2,m_\pi^2\right)}{2 m\left(\Xi^- \pi^+\right)} ,\nonumber  \\
        \lvert \vb{p}_- \rvert & = & \frac{\lambda^{1/2}\left(m^2\left(\Xi^- \pi^+\right),m_\Omega^2,m_\pi^2\right)}{2 m\left(\Xi^- \pi^+\right)} ,        
    \end{eqnarray}
are energies and momenta in the frame where $\vb{p}_\Xi + \vb{p}_+ = \vb{0}$. Here, $\lambda\left(a,b,c\right)$ is the Källén function, defined as
    \begin{equation}
        \lambda\left(a,b,c\right) = a^2 + b^2 + c^2 - 2\left(ab + bc + ca \right).
    \end{equation}

Let us estimate the LECs. There are several conventions for the LECs in use in the literature. In appendix \ref{app:convention} we give an overview. The values given here are used throughout the paper. Standard values are $D = 0.80$ and $F = 0.46$ \cite{Ledwig:2014rfa}. The parameter $h_\pi$ is related to kaon decays \cite{Jenkins:1991bt,Antipin:2007ya,Chivukula:1988gp,Cirigliano:2011ny} and is $h_\pi = -3.12 \times 10^{-7}$. The value for $h_A$ is determined from fits to different two-body decay channels of decuplet baryons \cite{Holmberg:2019ltw}. It lies within the range $1.9 \leq h_A \leq 2.9$. The value derived solely from cascade decays is $h_A = 2.0$. There are no simple observables to pin down $H_A$. To that end, we average estimates from large-$N_C$ \cite{Semke:2005sn,Pascalutsa:2006up,Ledwig:2011cx,Dashen:1993as} considerations and use $H_A = 2.0$. For a given $h_A$ we determine the magnitude of $h_C$ by fitting to the decay $\Omega^- \to \Xi^0 \pi^-$, whose decay width is proportional to $h_A^2 h_C^2$; cf. Eq.~\eqref{eq:Om-Xipi-both}. Therefore, $h_C$ lies between $3.5\times 10^{-8} \text{ GeV} \leq \lvert h_C \rvert \leq 5.4\times 10^{-8} \text{ GeV}$. From $h_A = 2.0$ we get $\lvert h_C \rvert = 5.1 \times 10^{-8}$ GeV. 

With the LECs fixed, we can calculate the double-differential decay width, given by \cite{pdg}
\begin{equation}
    \frac{{\rm d}^2\Gamma( \Omega^- \to \Xi^- \pi^+ \pi^-)}{{\rm d}m^2 \left(\Xi^- \pi^+ \right) {\rm d}m^2 \left(\Xi^- \pi^- \right)} = \frac{\langle \lvert \mathcal{M} \rvert^2 \rangle }{32 \left(2\pi m_\Omega \right)^3} ,
    \label{eq:15}
\end{equation}
where $\langle \lvert \mathcal{M} \rvert^2 \rangle$ denotes the spin-averaged Feynman matrix element. An explicit form of the Feynman matrix element is given in appendix \ref{app:MELform}. In appendix \ref{app:vertex} we provide some technical notes on deriving interaction vertices. When calculating the decay width we use \textsc{Mathematica} \cite{Mathematica} and \textsc{FeynCalc} \cite{Mertig:1990an,Shtabovenko:2016sxi,Shtabovenko:2020gxv} to take traces. 

\begin{figure*}[t] 
    \centering
    \begin{subfigure}{0.45\linewidth}
        \centering
        \includegraphics[width=.85\linewidth]{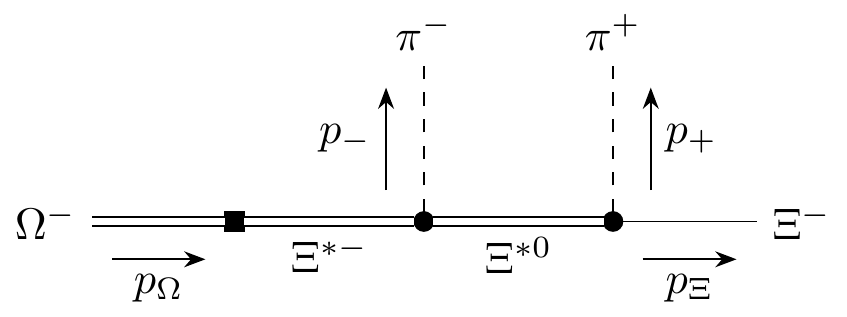}
        \caption{Diagram 1} 
    \end{subfigure}\quad
    \begin{subfigure}{0.45\linewidth}
        \centering
        \includegraphics[width=.85\linewidth]{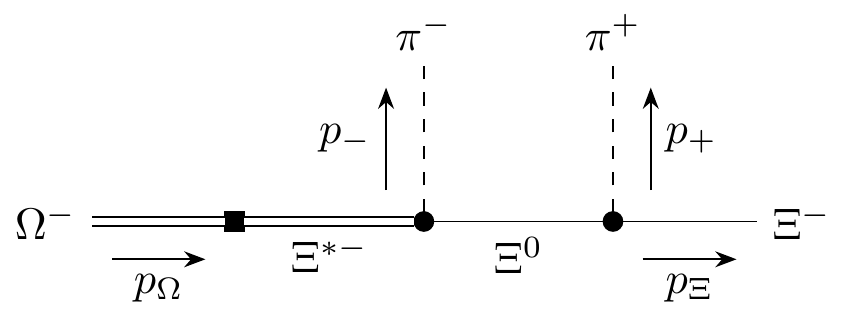} 
        \caption{Diagram 2} 
    \end{subfigure} 
    \begin{subfigure}{0.40\linewidth}
        \centering
        \includegraphics[width=.75\linewidth]{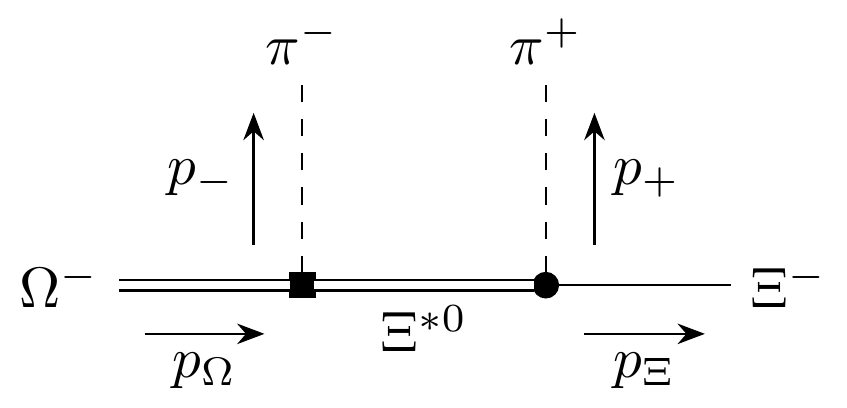} 
        \caption{Diagram 3} 
    \end{subfigure} \quad
    \begin{subfigure}{0.43\linewidth}
        \centering
        \includegraphics[width=.50\linewidth]{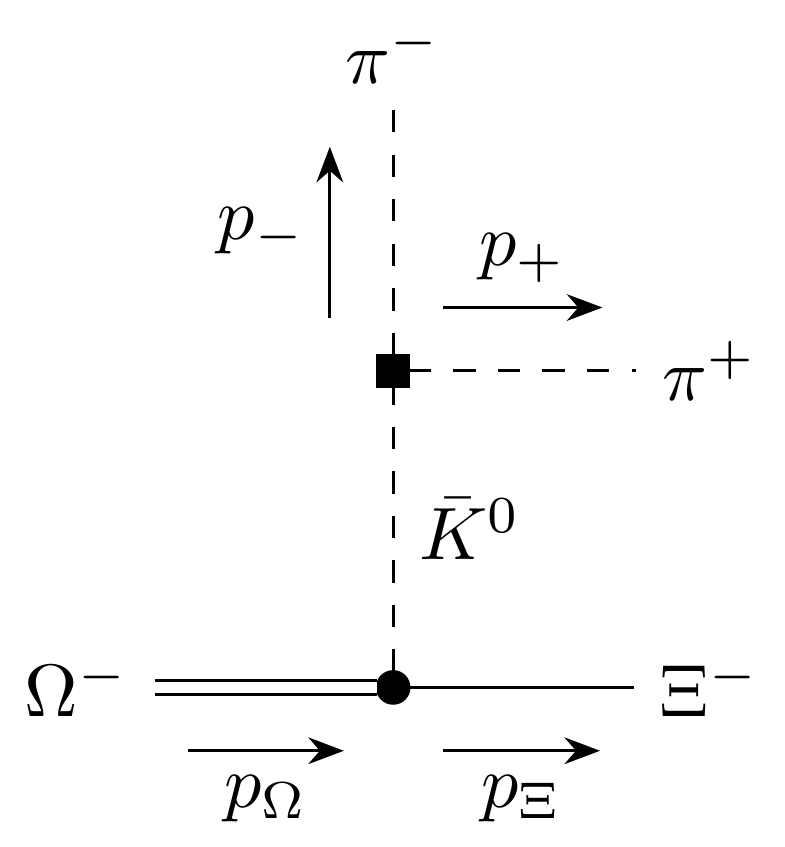} 
        \caption{Diagram 4} 
    \end{subfigure}
    \caption{Diagrams contributing to the decay $\Omega^- \to \Xi^- \pi^+ \pi^-$. Circles indicate a strong vertex and squares indicate a weak vertex.}
    \label{fig:feynd}
\end{figure*}

\begin{figure}[t]
    \centering
    \includegraphics[width=1\linewidth]{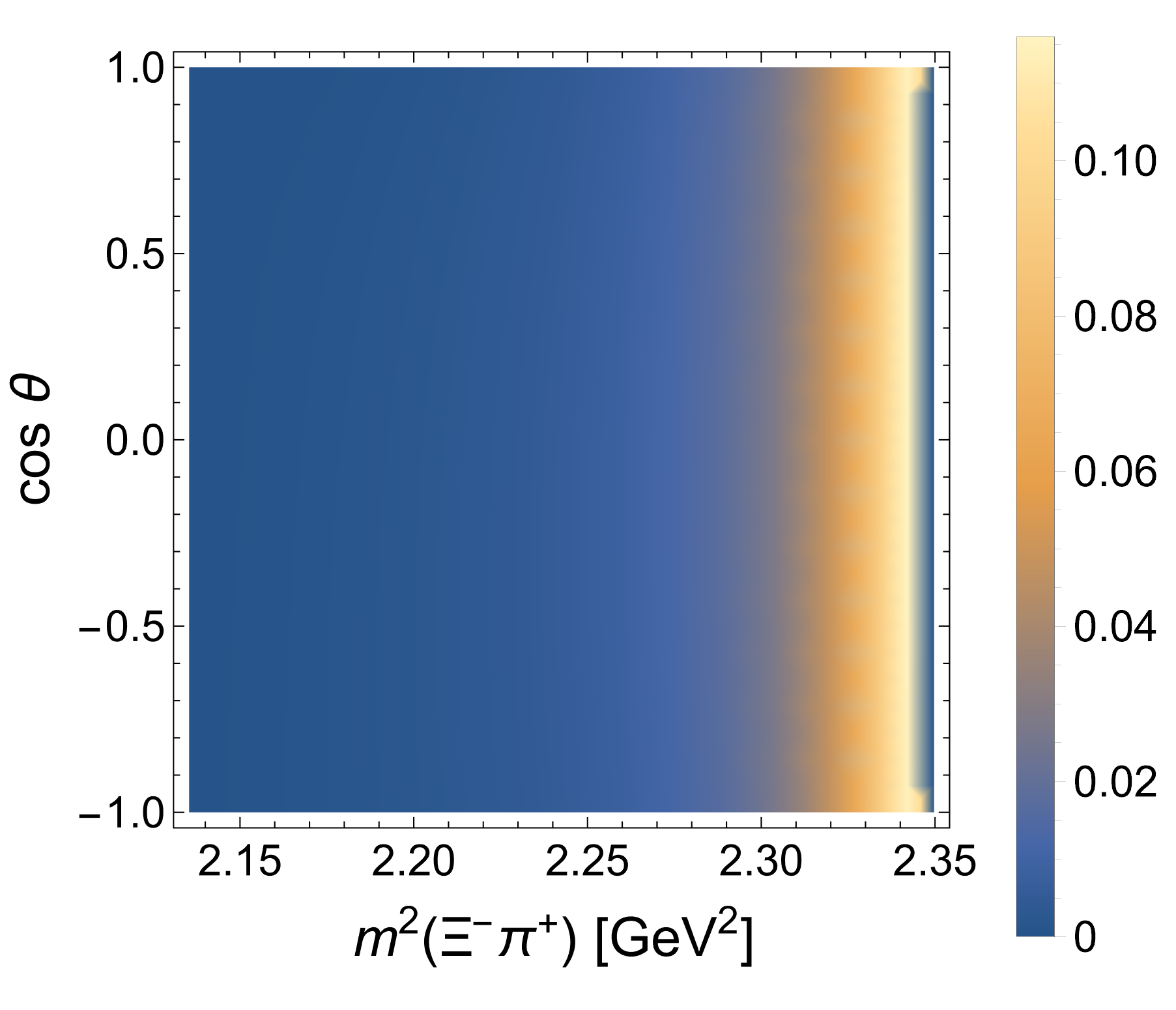}
    \caption{The fully-differential distribution for the decay $\Omega \to \Xi^- \pi^+ \pi^-$ at leading order, normalized to the total decay width. The $\Xi\left(1530\right)$ resonance overwhelms all other features of the distribution.}
    \label{fig:dalcomplete}
\end{figure}

As can be seen in Fig.~\ref{fig:feynd}, the LO Lagrangians \eqref{eq:strong}, \eqref{eq:strong-meson}, \eqref{eq:weak-mb} provide four tree-level diagrams contributing to the decay. In line with the discussions in Subsec.\ \ref{sec:Xipi}, the numerically dominant term of the decay width comes from diagram 3 and is proportional to $h_A^2 h_C^2$. In Fig.~\ref{fig:dalcomplete} we see it is the well-known $\Xi\left(1530\right)$ resonance (see also Ref.~\cite{Antipin:2007ya}), which completely overwhelms all other features in the phase space. This matches with what is known from heavy-baryon calculations \cite{Antipin:2007ya}. Relativistic corrections to the branching fraction are on the order of ten percent as compared to the heavy-baryon results \cite{Mommers1677385}.

Our results do not match with what has been seen in the most recent experiment \cite{HyperCP:2010ego}. There, no trace of the $\Xi(1530)$ resonance has been observed. Data seem to be compatible with statistical phase space. As discussed in Ref.~\cite{Antipin:2007ya}, the decay width is sensitive to the values of the LECs. Therefore, we resort to the ratio of three- and two-body decays to obtain a robust, model-independent result. We will see that the discrepancy is not resolved by tuning parameters. 

In the spin-averaged decay width, parity-conserving processes (diagrams 1-2) do not interfere with parity-violating processes (diagrams 3-4). To see this, we factor the decay width as
\begin{equation}\label{eq:decwidthdecomp}
    \Gamma \propto \tr\left\{ \left[ \left( \text{v type} \right) + \left( \text{c type} \right)\gamma_5  \right] \left[ \left( \text{v type} \right) + \left( \text{c type} \right)\gamma_5 \right] \right\}.
\end{equation}
Interference terms contain several gamma matrices and exactly one $\gamma_5$ matrix. It follows that the trace must vanish or be proportional to a Levi-Civita symbol. The Levi-Civita symbol must contract with the particle momenta. Since all momenta lie in the same plane this contraction gives zero. Therefore,
\begin{equation}\label{eq:Gammadecomp}
    \Gamma = \lvert \Gamma_c \rvert^2 + \lvert \Gamma_v \rvert^2 \geq \lvert \Gamma_v \rvert^2.
\end{equation}

When the LECs are varied within their physical limits, then we find in all cases that $\lvert \Gamma_c \rvert^2$ is numerically small as compared to $\lvert \Gamma_v \rvert^2$ (on the order of a couple of percent). Again, this supports our claim that the dominant contribution to the three-body decay relates to the two-body decay by chiral symmetry. 

We can expand the parity-violating part as 
\begin{eqnarray}
  \lvert \Gamma_v \rvert^2 &=& a\left(h_A h_C\right)^2 + b\left(h_A h_\pi\right)^2 + c h_A^2 h_C h_\pi   \nonumber \\
  &=& a\left(h_A h_C\right)^2 \left( 1 + \frac{b}{a} \frac{h_\pi^2}{h_C^2} + \frac{c}{a} \frac{h_\pi}{h_C} \right)  \,,
  \label{eq:17}
\end{eqnarray}
where $a$, $b$ and $c$ are numbers that only depend on the kinematic factors and the pion decay constant, 
but not on other coupling constants. The dependence on the parameter combinations follows from the formulae in 
Appendix \ref{app:MELform}. For realistic values of the parameters $h_\pi$ and $h_C$ the last 
two terms $\sim b/a$ and $\sim c/a$ stay rather small (on the order of $10\%$). 
The dominant term, on the other hand, depends on the same parameter combination as the two-body decay 
amplitude $\Omega^- \to \Xi \pi$. 

This allows us to provide a robust and parameter insensitive estimate for the lower bound of the ratio of the branching fractions of the two-and-three-body decays,
\begin{equation}
    \frac{\mathcal{B}\left(\Omega^- \to \Xi^-\pi^+\pi^-\right)}{\mathcal{B}\left(\Omega^- \to \Xi^0 \pi^- \right) }  \gtrsim 
    \frac{\lvert \Gamma_v \rvert^2}{\Gamma(\Omega^- \to \Xi^0 \pi^-)} \,.
    \label{eq:18}
\end{equation} 
For the numerical estimate we vary the parameter $h_C$ in a reasonable range. 
Recall, however, that our fit for $h_C$ only fixes its magnitude and not its sign. So, we calculate the lower bound for both signs of $h_C$ and pick whichever case gives the lowest bound (this automatically covers the other sign choice as well). We find
\begin{equation}
     \frac{\mathcal{B}\left(\Omega^- \to \Xi^-\pi^+\pi^-\right)}{\mathcal{B}\left(\Omega^- \to \Xi^0 \pi^- \right) } \gtrsim 2.9 \times 10^{-2}.
    \label{eq:19}
\end{equation}
If we drop the terms $\sim b/a$ and $\sim c/a$ we obtain 
\begin{equation}
     \frac{\mathcal{B}\left(\Omega^- \to \Xi^-\pi^+\pi^-\right)}{\mathcal{B}\left(\Omega^- \to \Xi^0 \pi^- \right) } \gtrsim 3.1 \times 10^{-2}  \,,
    \label{eq:19a}
\end{equation}
fairly close to Eq.~\eqref{eq:19}. 

If we use the measured branching fraction of the decay $\Omega^- \to \Xi^0 \pi^+ \pi^-$ \cite{HyperCP:2010ego} we get
\begin{equation}
    \frac{\mathcal{B}\left( \text{HyperCP} \right)}{\mathcal{B}\left(\Omega^- \to \Xi^0 \pi^- \right) } = \left( 0.16 \pm 0.03 \right) \times 10^{-2}.
    \label{eq:20}
\end{equation}
There is an order-of-magnitude difference between theory and experiment. We stress again the intimate link between this discrepancy and a possible violation of the $\Delta I = 1/2$ rule. Obviously, the three-body decay is much better suited to explore this discrepancy, as the order-of magnitude-difference for the three-body decay is much larger as compared to the discrepancy for the two-body decays expressed by the difference between ratios \eqref{eq:Omega-DeltaI12} and \eqref{eq:Omega-DeltaI12-exp}. 

At present though, it is unclear whether the resolution of this tension should come from improvements on the theory side, experimental side, or both. Within the heavy-baryon formalism the inclusion of higher-order terms and isospin-3/2 contributions has yielded inconclusive results \cite{Antipin:2007ya,Tandean:1998ch}. Empirical data are scarce. Simply put, we urge that the decay $\Omega^- \to \Xi^- \pi^+ \pi^-$ be remeasured at ongoing and upcoming experiments, such as BESIII, LHCb, Belle-II, and PANDA. Ideally, fully differential data will match more closely to current results, or trigger further improvements on the theory side. 

\subsection{Other $\Omega^- \to \Xi \pi \pi$ decays}

We can extend our reasoning to the so-far unmeasured decays $\Omega^- \to \Xi^0 \pi^- \pi^0$ and $\Omega^- \to \Xi^- \pi^0 \pi^0$. The diagrams look like Fig.~\ref{fig:feynd}, with obvious particle replacements. There are three additional diagrams similar to diagrams 1-3 where the pions are swapped. Hence, the decay width has the same structure as Eq.~$\left(\ref{eq:Gammadecomp}\right)$ and we can make predictions for the lower bound of the ratio of branching fractions. We find
\begin{eqnarray}
  \frac{\mathcal{B}\left(\Omega^- \to \Xi^0\pi^-\pi^0\right)}{\mathcal{B}\left(\Omega^- \to \Xi^0 \pi^- \right) } &\gtrsim& 5.1 \times 10^{-2} ,  \nonumber \\*
  \frac{\mathcal{B}\left(\Omega^- \to \Xi^-\pi^0\pi^0\right)}{\mathcal{B}\left(\Omega^- \to \Xi^0 \pi^- \right) } &\gtrsim& 1.2 \times 10^{-2}.
  \label{eq:100}
\end{eqnarray}  
That the former ratio is similar in size to the ratio of the decay $\Omega^- \to \Xi^- \pi^+ \pi^-$ while the latter ratio is smaller follows from the relative sizes of the Clebsch-Gordan coefficients corresponding to the resonance vertices.

\begin{figure*}[t] 
    \centering
    \begin{subfigure}{0.45\linewidth}
        \centering
        \includegraphics[width=1\linewidth]{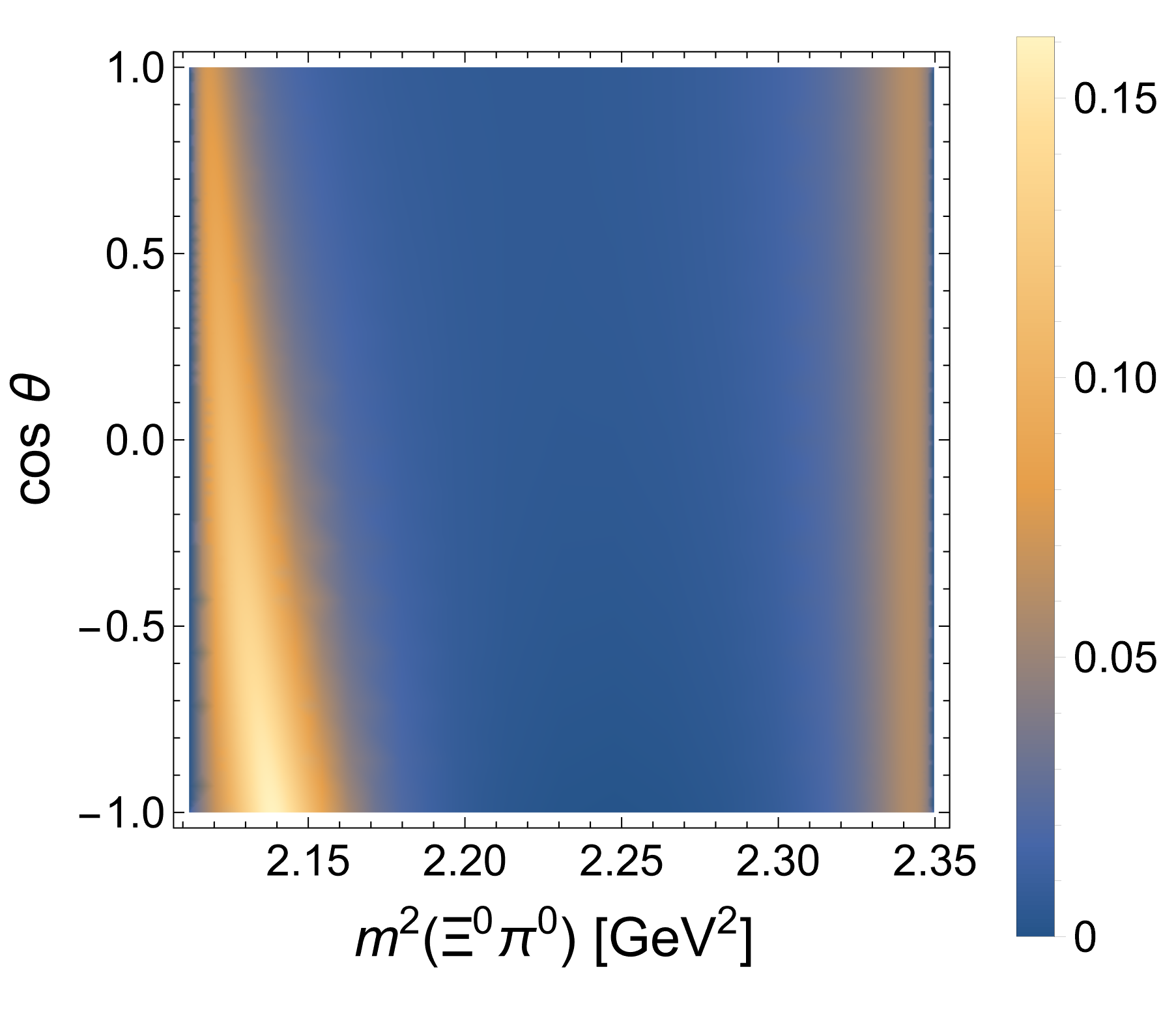}
        \caption{Fully-differential distribution for the decay $\Omega^- \to \Xi^0 \pi^- \pi^0$.} 
    \end{subfigure}\quad
    \begin{subfigure}{0.45\linewidth}
        \centering
        \includegraphics[width=1\linewidth]{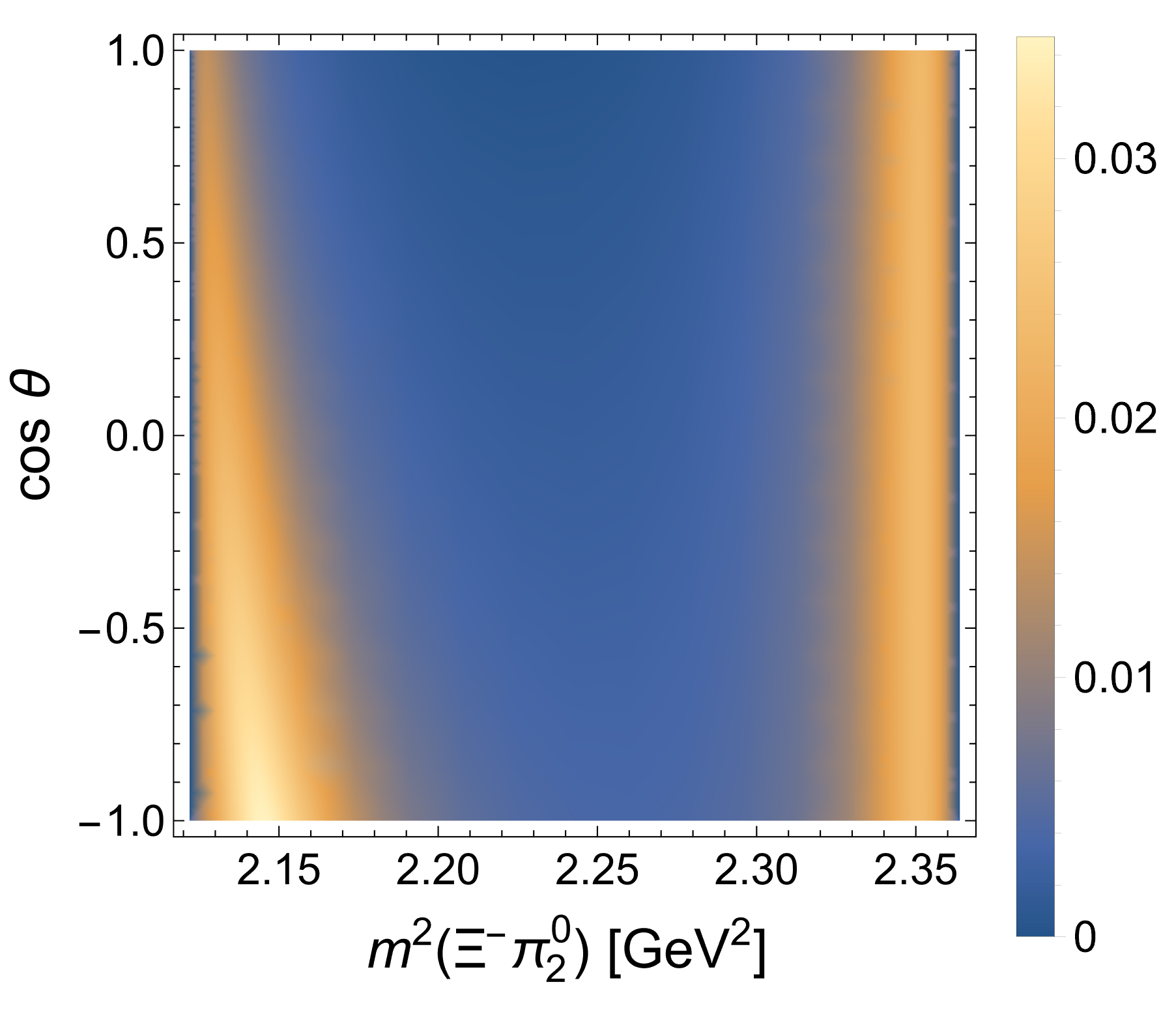} 
        \caption{Fully-differential distribution for the decay $\Omega^- \to \Xi^- \pi^0 \pi^0$.} 
    \end{subfigure} 
    \caption{Fully-differential distributions for three-body Omega decays, normalized by the total decay width. Both distributions are double peaked. The right peak is the $\Xi\left(1530\right)$ resonance. The left peak arises due to symmetry reasons and is slightly smeared out, as can be seen in both figures.}
    \label{fig:dal23}
\end{figure*}

For completeness, let us also look at the fully-differential distributions. They are given in Fig.~\ref{fig:dal23}. For the decay $\Omega^- \to \Xi^0 \pi^- \pi^0$ we use $m^2\left(\Xi^0 \pi^0\right) = \left(p_\Xi + p_0 \right)^2$ and the angle $\theta$ between $\vb{p}_\Xi$ and $\vb{p}_-$ in the frame where $\vb{p}_\Xi + \vb{p}_0 = \vb{0}$. For the decay $\Omega^- \to \Xi^- \pi^0 \pi^0$ we use $m^2\left(\Xi^- \pi^0_2\right) = \left(p_\Xi + p_{02}\right)^2$ and the angle between $\vb{p}_\Xi$ and $\vb{p}_{01}$ in the frame where $\vb{p}_\Xi + \vb{p}_{02} = \vb{0}$. Here, we label the two neutral pions with 1 and 2.

Both distributions are double-peaked, which is not entirely surprising. The right peak is the $\Xi\left(1530\right)$ resonance. The left peak arises due to symmetry reasons. The three new diagrams that our decays have compared to the decay $\Omega^- \to \Xi^- \pi^+ \pi^-$ are symmetric under exchange of the pion momenta. This means that the fully-differential distribution in the two kinematic variables should be symmetric as well. When we change one of the kinematic variables to an angle the left peak gets smeared out. This is also visible in the distributions.  

Another hard prediction, besides the lower bounds \eqref{eq:19} and \eqref{eq:100}, is that the resonance peaks should be clearly visible in the fully-differential distributions displayed in 
Figs.~\ref{fig:dalcomplete} and \ref{fig:dal23}. In turn, high-quality differential distributions can be used to study the relative importance of the $\Delta I=1/2$ and $\Delta I = 3/2$ amplitudes. The corresponding theoretical study requires such an experimental input and is therefore beyond the scope of the present work.

\section{Semileptonic decays}
\label{sec:munu}

While there are numerous LECs in the weak NLO Lagrangian \cite{Borasoy:1998ku} 
that provides corrections to Eq.~\eqref{eq:weak-mb}, the ``strong'' NLO Lagrangian has been fully specified 
in Ref.~\cite{Holmberg:2018dtv} 
for the strong-interaction sector plus its coupling to external fields. We will use it to provide predictions for the 
decay $\Omega^- \to \Xi^0 \mu^- \bar{\nu}_\mu$. 

Before turning to the technical aspects, we have to clarify what ``NLO'' means in the present context. In general, we study 
baryon decays of type $B_1 \to B_2 + X$ where $X$, denotes one or several mesons, photons, or dileptons. Parametrically, the 
baryon mass is large and the mass difference between the baryons is small. The masses of all particles contained in $X$ are small.
Consequently, in the rest frame of the decaying state $B_1$ the modulus of the momenta of $B_2$ and $X$ is small. This qualitative statement about the smallness of momenta does not change if we change to the rest frame of $B_2$. 

But how small are such momenta? Kinematics tells 
\begin{eqnarray}
 \vert \vb{p}_2 \vert & = & \vert \vb{p}_X \vert \nonumber \\
  & = & \frac{1}{2 m_1} \left[ 
    \left((m_1+m_2)^2-m_X^2 \right) \left((m_1-m_2)^2-m_X^2 \right)  \right]^{1/2} \nonumber \\
  & \approx & 
 \left[ (m_1-m_2)^2-m_X^2 \right]^{1/2},
  \label{eq:pcm-b1rest}
\end{eqnarray}
which holds in the frame where $B_1$ is at rest. Here, $m_i$ ($\vb{p}_i$) is the mass (three-momentum) 
of baryon $i$ or of the single- or many-particle state $X$. 

Before turning to our case, a small digression is in order to understand the overall picture. If {\em both} 
$B_1$ and $B_2$ belong to the lowest-lying baryon octet \eqref{eq:3}, then their mass difference is driven by the masses of the 
light quarks. Given that the strange-quark mass is significantly larger than the masses of up and down quark, we can 
approximate $m_2 - m_1 \sim m_s \sim M_K^2$, as pointed out in Ref.~\cite{Jenkins:1991bt}. For kinematic reasons, the final state 
$X$ cannot contain kaons ($m_1-m_2 < M_K$). 
One can assume that $m_X$ is parametrically smaller than $M_K$. Either it is $\sim M_K^2$ or suppressed even 
further. This leads to $\vert \vb{p}_2 \vert = \vert \vb{p}_X \vert \sim m_1 - m_2 \sim M_K^2$. In contrast to the 
typical situation studied in meson-baryon scattering, the external momenta are not as large as the Goldstone-boson mass, 
here $M_K$, but further suppressed. Derivative counting in the chiral Lagrangian might not go along with kaon-mass counting. 
For such hyperon decays, loop momenta count as $M_K$ while external momenta count as $M_K^2$ \cite{Jenkins:1991bt}.

But, there is a twist for the decuplet. The main decay branch of the $\Omega^-$ baryon is $\Omega^- \to \Lambda K^-$. If we put 
$M_X = M_K$ in Eq.~\eqref{eq:pcm-b1rest}, we obtain the ``standard'' scaling 
$\vert \vb{p}_\Lambda \vert = \vert \vb{p}_K \vert \sim M_K$ -- provided we use $m_\Omega - m_\Lambda \sim M_K$. 
Definitely, we cannot use $m_\Omega - m_\Lambda \sim M_K^2$. This would lead to a contradiction (imaginary momentum) 
in Eq.~\eqref{eq:pcm-b1rest}. For all other decays of the $\Omega^-$ baryon there is no predefined answer about the size of the 
baryon mass difference relative to the Goldstone boson mass(es). For the present work we have decided to deal with all $\Omega^-$ 
decays on equal footing. Therefore we assign \cite{Holmberg:2018dtv,Holmberg:2019ltw} 
\begin{eqnarray}
  \label{eq:our-power-counting}
  m_{\rm decuplet} - m_{\rm octet} \sim M_K, 
\end{eqnarray}
and as a consequence $\vert \vb{p}_2 \vert = \vert \vb{p}_X \vert \sim m_{\rm decuplet} - m_{\rm octet} \sim M_K$. All soft derivatives 
appearing in Eq.~\eqref{eq:strong} and Eq.~\eqref{eq:strong-NLO-Lagr} count as order ${\cal O}(M_K)$. This is in full analogy to the 
small-scale expansion $m_\Delta - m_N \sim M_\pi$ in the two-flavor sector \cite{Hemmert:1997ye}. 
With this power counting, the NLO calculation of the decay $\Omega^- \to \Xi^0 \mu^- \bar{\nu}_\mu$ is a tree-level calculation.
Loops start to contribute only at NNLO. 

\begin{figure*}[t] 
    \centering
    \begin{subfigure}{0.45\linewidth}
        \centering
        \includegraphics[width=.7\linewidth]{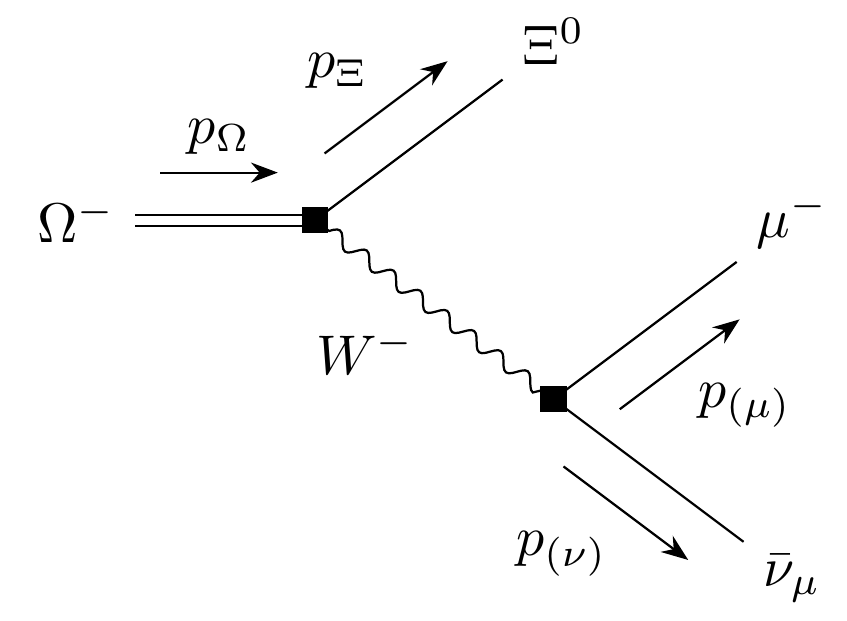}
        \caption{Diagram 1} 
    \end{subfigure}\quad
    \begin{subfigure}{0.45\linewidth}
        \centering
        \includegraphics[width=.9\linewidth]{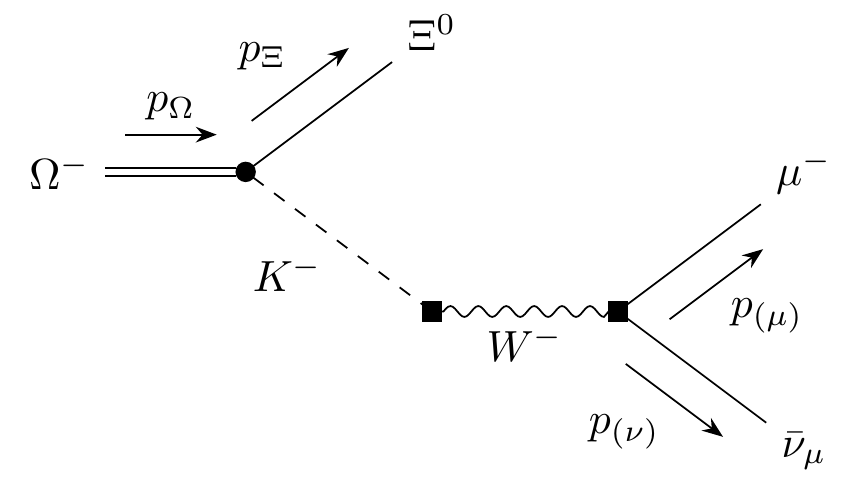} 
        \caption{Diagram 2} 
    \end{subfigure} 
    \caption{LO and NLO diagrams contributing to the decay $\Omega^- \to \Xi^0 \mu^- \bar{\nu}_{\mu}$,
      see also \eqref{eq:34}, \eqref{eq:35}.}
    \label{fig:NLOsemilep}
\end{figure*}

Now, we turn to the technical aspects of the decay $\Omega^- \to \Xi^0 \mu^- \bar{\nu}_\mu$. The diagrams contributing at NLO are given in Fig.~\ref{fig:NLOsemilep}. Our kinematic variables are $m^2\left(\mu^- \bar{\nu}_\mu\right) = \left(p_{\left(\mu\right)} + p_{\left(\nu\right)} \right)^2$ and $\cos\theta$, where $\theta$ is the angle between $\vb{p}_{\left(\nu\right)}$ and $\vb{p}_\Xi$ in the frame where $\vb{p}_{\left(\mu\right)} + \vb{p}_{\left(\nu\right)} = \vb{0}$.

\begin{figure}[t] 
    \centering
    \begin{subfigure}{0.45\linewidth}
        \centering
        \includegraphics[width=1\linewidth]{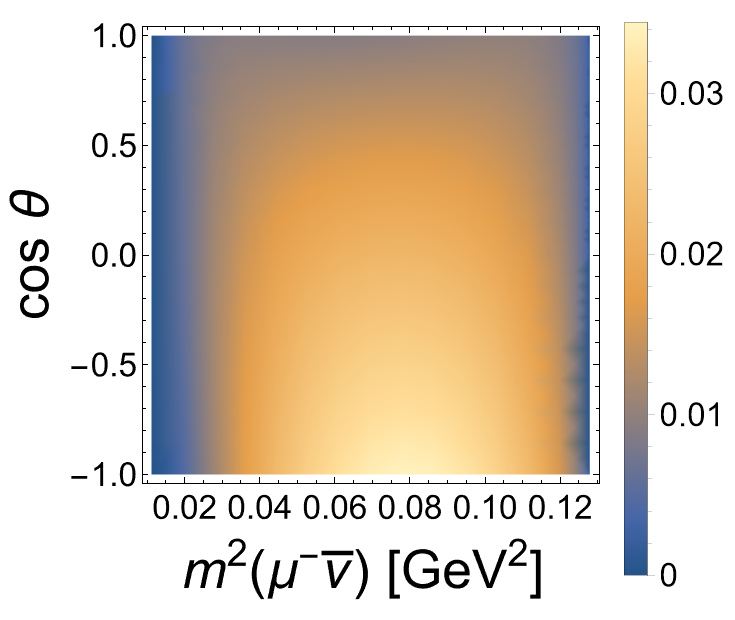}
        \caption{$c_M = 1.92 \text{ GeV}^{-1}$, $c_E = 0.5 \text{ GeV}^{-1}$}\label{fig:dala} 
    \end{subfigure}\quad
    \begin{subfigure}{0.45\linewidth}
        \centering
        \includegraphics[width=1\linewidth]{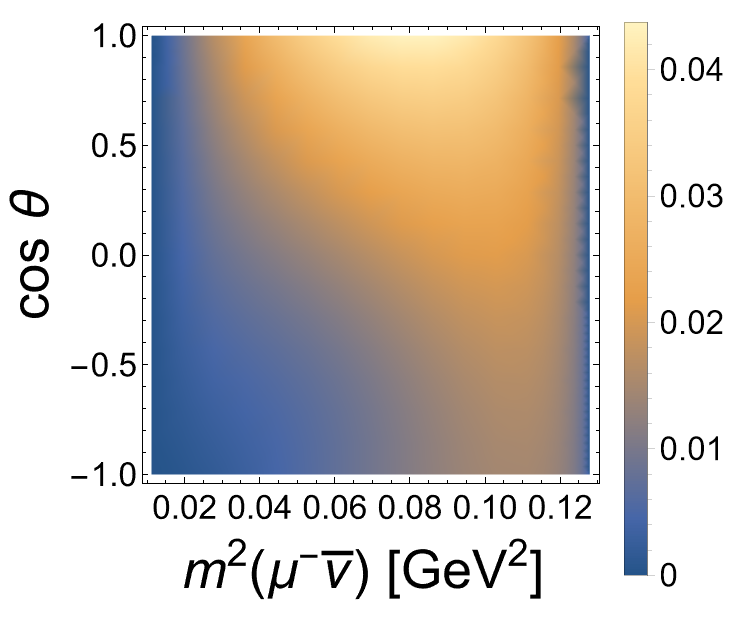} 
        \caption{$c_M = 1.92 \text{ GeV}^{-1}$, $c_E = -5 \text{ GeV}^{-1}$}\label{fig:dalb}
    \end{subfigure} 
    \begin{subfigure}{0.45\linewidth}
        \centering
        \includegraphics[width=1\linewidth]{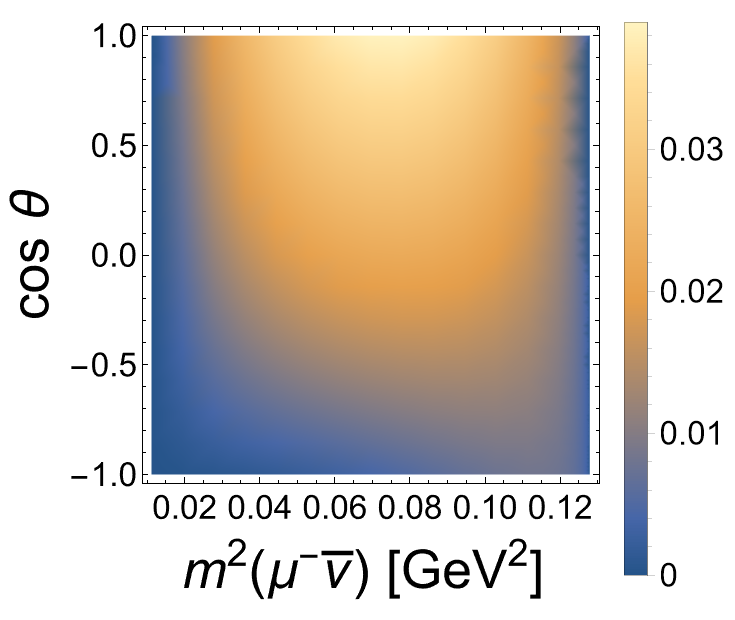} 
        \caption{$c_M = -1.92 \text{ GeV}^{-1}$, $c_E = 0.5 \text{ GeV}^{-1}$}\label{fig:dalc} 
    \end{subfigure} \quad
    \begin{subfigure}{0.45\linewidth}
        \centering
        \includegraphics[width=1\linewidth]{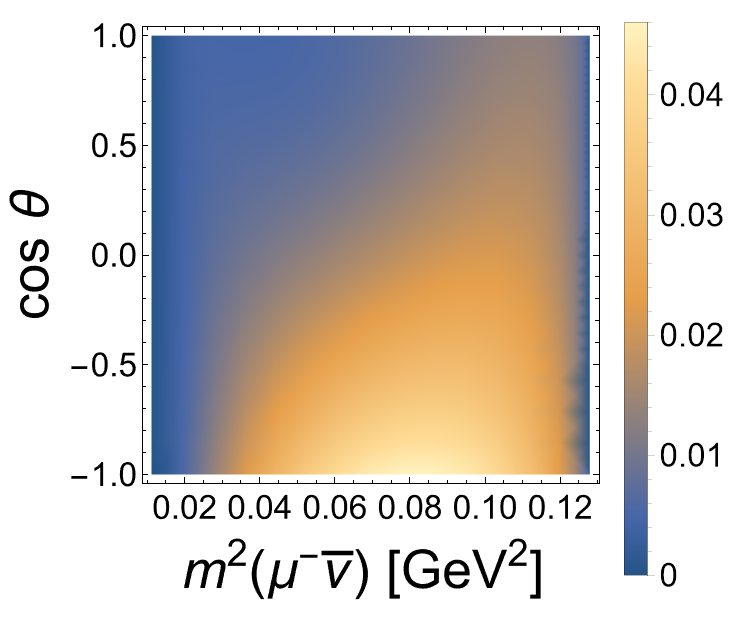} 
        \caption{$c_M = -1.92 \text{ GeV}^{-1}$, $c_E = -5 \text{ GeV}^{-1}$}\label{fig:dald} 
    \end{subfigure}
    \caption{Fully-differential distribution for the branching ratio of the decay $\Omega^- \to \Xi^0 \mu^- \bar{\nu}_\mu$ at NLO for different values of the LECs $c_E$ and $c_M$. Plots are normalized by the total decay width of the Omega baryon.}
    \label{fig:dalNLO}
\end{figure}

\begin{figure}[t]
    \centering
    \includegraphics[width=1\linewidth]{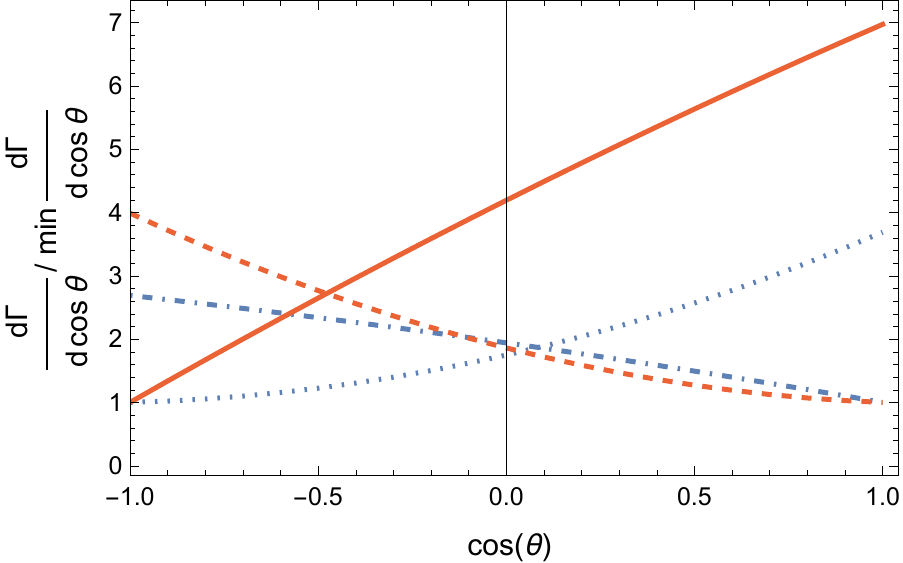}
    \caption{The normalized angular distribution for the decay $\Omega^- \to \Xi^0 \mu^- \bar{\nu}_\mu$ at NLO. The dash-dotted blue line corresponds to the parameter values in fig.~\ref{fig:dala}, the dotted blue line corresponds to fig.~\ref{fig:dalb}, the solid orange line corresponds to fig.~\ref{fig:dalc} and the dashed orange line corresponds to fig.~\ref{fig:dald}.}
    \label{fig:ang}
\end{figure}

At NLO the decay width depends on two new LECs, $c_E$ and $c_M$. Our proposal is that they can be pinned down by an experimental
determination of the differential decay width. To provide some concrete estimates 
we use Refs.~\cite{Holmberg:2018dtv, Holmberg:2019ltw}, where the magnitude of each of these constants has been narrowed down 
to one of two possible values: 
\begin{equation*}
    c_E = 0.5 \text{ GeV}^{-1} \text{ or } -5 \text{ GeV}^{-1} \quad c_M = \pm 1.92 \text{ GeV}^{-1}. 
\end{equation*}
As can be seen in Fig.~\ref{fig:dalNLO}, it is possible to distinguish the four configurations of the LECs. It is also possible to distinguish all four cases just by looking at the angular distributions, as can be seen in Fig.~\ref{fig:ang}. However, if the data are integrated over $\cos \theta$ it is only possible to pin down the relative sign of $c_E$ and $c_M$. 

Like the analysis done in Ref.~\cite{Holmberg:2019ltw} for the decay $\Omega^- \to \Xi^0 e^- \bar{\nu}_e$ \footnote{In passing we remark that the fully-differential distribution in Ref.~\cite{Holmberg:2019ltw} appears to be flipped with respect to ours. This is because the authors use the angle between the cascade and the electron, while we use the angle between the cascade and the neutrino.}, it appears that $c_M \mapsto -c_M$ is essentially equivalent to $\cos\theta \mapsto -\cos\theta$ (forward-backward asymmetry \cite{Donoghue:1992dd}). This is not exactly true. Closer inspection of, for instance, Fig.~\ref{fig:dala} and Fig.~\ref{fig:dalc} reveals that there is a degree of asymmetry in this equivalence. This asymmetry arises because the leptons have finite mass. Compared to the electronic case, this asymmetry is more pronounced because muons are heavier than electrons. 

When we use $c_M = 1.92$ GeV$^{-1}$ the predicted branching fraction is lower by approximately $30 \%$ as compared to the electronic case. As a function of $c_E$ it has a similar parabolic shape as in Ref.~\cite{Holmberg:2019ltw}. Its value is $\mathcal{B}\left(\Omega^- \to \Xi^0 \mu^- \bar{\nu}_\mu\right) = 3.79 \times 10^{-3} $ for $c_E = 0.5\,$GeV$^{-1}$.

\section{Outlook}
\label{sec:outlook}

For the decay $\Omega^- \to \Xi^- \pi^+ \pi^-$ we have demonstrated an order-of-magnitude discrepancy between experiment, Eq.~\eqref{eq:20}, and theory, Eq.~\eqref{eq:19}, with the latter based on the $\Delta I = 1/2$ selection rule. This adds to the well-known discrepancy for the two-body decays expressed in Eqs.~\eqref{eq:Omega-DeltaI12} and \eqref{eq:Omega-DeltaI12-exp}. Therefore, we recommend the decays $\Omega^- \to \Xi^- \pi^+ \pi^-$ as well as $\Omega^- \to \Xi^- \pi^0$ and $\Omega^- \to \Xi^0 \pi^-$ be remeasured at experiments such as BESIII, LHCb, Belle-II, and PANDA. 
Besides these branching fractions the related interesting question is the appearance or non-appearance of the 
$\Xi(1530)$ resonance peak(s) in the fully-differential distributions of the decays $\Omega^- \to \Xi \pi \pi$. If the processes are dominated 
by the $\Delta I = 1/2$ selection rule, this resonance should be visible. Therefore, differential distributions, even if not normalized, would be highly welcome to understand the fate of the intermediate $\Xi(1530)$ state. 
We have also calculated lower limits for the branching fraction of the decays $\Omega^- \to \Xi^0 \pi^- \pi^0$ and $\Omega^- \to \Xi^- \pi^0 \pi^0$. Lastly, we have looked at the decay $\Omega^- \to \Xi^0 \mu^- \bar{\nu}_\mu$ and shown that measurements thereof may be used to pin down the low-energy constants $c_M$ and $c_E$.

\section*{Acknowledgments}
SL thanks L.\ Eklund and A.\ Kup\'s\'c for inspiring discussions. CJGM wishes to thank D.\ An, N.~Salone and F.\ Alvarado for useful talks.
This work has been supported by the Swedish Research Council (Vetenskapsr\aa det) (grant number 2019-04303).

\appendix

\section{Explicit matrix-element formulae}
\label{app:MELform}

For the decays $\Omega^- \to \Xi \pi \pi$ the generic structure of the Feynman matrix element is
\begin{equation}
    \mathcal{M}_i(\Omega^- \to \Xi \pi \pi) = \bar{u}( p_\Xi) \, \beta_i^\mu \, u_\mu(p_\Omega),
    \label{eq:22}
\end{equation}
where $\beta_i^\mu$ depends on what decay $i$ we consider. Here, $u_\mu(p,\sigma)$ are vector-spinors \cite{Rarita:1941mf} describing spin-3/2 states with momentum $p$ and polarization $\sigma$. They satisfy
\begin{equation}
    \sum_\sigma u_\mu(p,\sigma) \, \bar{u}_\nu (p,\sigma) = -\left(\slashed{p} + m \right) P^{3/2}_{\mu\nu}(p),
    \label{eq:23}
\end{equation}
with $p^0 = \sqrt{m^2 + \vb{p}^2}$ the energy of the spin-3/2 particle and $m$ its mass. The projector on the spin-3/2 subspace \cite{deJong:1992wm} is defined as
\begin{equation}
    P^{3/2}_{\mu\nu}(p) = g_{\mu\nu} - \frac{1}{3}\gamma_\mu \gamma_\nu - \frac{1}{3p^2}\left(\slashed{p}\gamma_\mu p_\nu + p_\mu \gamma_\nu \slashed{p}\right).
    \label{eq:24}
\end{equation}

For decay 1, $\Omega^- \to \Xi^- \pi^+ \pi^-$, we find
\begin{eqnarray}
    \beta^\sigma_1 &&= \frac{h_A H_A h_C}{18 \sqrt{2} F_\pi^2} p_+^\mu P^T_{\mu\nu}(k_1, m_{\Xi^\ast}, \Gamma_{\Xi^\ast} ) \nonumber \\*
    &&{} \times \slashed{p}_- \gamma_5 P_T^{\nu\sigma}(p_\Omega,m_{\Xi^\ast}) \nonumber \\*
    &&{} + \frac{h_A h_C \left(D - F \right)}{6 \sqrt{2} F_\pi^2} p^-_{\mu} \slashed{p}_+ \gamma_5 \nonumber \\*
    &&{} \times P^B(k_1,m_\Xi ) \, P_T^{\mu\sigma}(p_\Omega, m_{\Xi^\ast}) \nonumber \\*
    &&{} -\frac{i h_A h_C}{6\sqrt{2} F_\pi^2} p^+_{\mu} P_T^{\mu\sigma}(k_1, m_{\Xi^\ast}, \Gamma_{\Xi^\ast}) \nonumber \\*
    &&{} - \frac{i h_A h_\pi}{4\sqrt{2} F_\pi^2} P^\Phi(\tilde{k}, m_K) \, \tilde{k}^\sigma \left( \tilde{k} \cdot p_+ + p_+ \cdot p_- \right). \nonumber \\ &&
    \label{eq:30}
\end{eqnarray}
For decay 2, $\Omega^- \to \Xi^0 \pi^- \pi^0$, we find
\begin{eqnarray}
    \beta^\sigma_2 &&= \bigg[ \frac{h_AH_Ah_C}{36 F_\pi^2} p_0^\mu P_{\mu\nu}^T(k_1,m_{\Xi^\ast},\Gamma_{\Xi^\ast}) \nonumber \\*
    &&{} \times \slashed{p}_- \gamma_5 P_T^{\nu \sigma}(p_\Omega,m_{\Xi^\ast}) \nonumber \\*
    &&{} + \frac{h_A h_C \left(D-F\right)}{12 F_\pi^2} p^-_\mu \slashed{p}_0 \gamma_5 \nonumber \\*
    &&{} \times  P^B(k_1,m_\Xi) \, P_T^{\mu\sigma}(p_\Omega, m_{\Xi^\ast})    \nonumber \\*
    &&{} + \frac{ih_A h_C}{12 F_\pi^2}  p^0_\mu P_T^{\mu\sigma}(k_1, m_{\Xi^\ast}, \Gamma_{\Xi^\ast}) - \left(p_- \leftrightarrow p_0 \right)  \bigg] \nonumber \\*
    &&{} + \frac{ih_A h_\pi}{8 F_\pi^2} P^\Phi(\tilde{k},m_K) \, \tilde{k}^\sigma \left(\tilde{k}\cdot p_- - \tilde{k} \cdot p_0\right).
    \label{eq:31}
\end{eqnarray}
For decay 3, $\Omega^- \to \Xi^- \pi^0 \pi^0$, we find
\begin{eqnarray}
    \beta^\sigma_3 &&= \bigg[ -\frac{h_AH_Ah_C}{36 \sqrt{2} F_\pi^2} p_{20}^\mu P_{\mu\nu}^T(k_1,m_{\Xi^\ast},\Gamma_{\Xi^\ast}) \nonumber \\*
    &&{} \times \slashed{p}_{10} \gamma_5 P_T^{\nu \sigma}(p_\Omega,m_{\Xi^\ast})  \nonumber\\*
    &&{} + \frac{h_A h_C \left(D-F\right)}{12 \sqrt{2} F_\pi^2}  p^{10}_\mu \slashed{p}_{20} \gamma_5 \nonumber \\*
    &&{} \times P^B(k_2,m_\Xi) \, P_T^{\mu\sigma}(p_\Omega, m_{\Xi^\ast})  \nonumber \\*
    &&{} + \frac{ih_A h_C}{12 \sqrt{2} F_\pi^2}  p^{20}_\mu P_T^{\mu\sigma}(k_2, m_{\Xi^\ast}, \Gamma_{\Xi^\ast}  ) + \left( p_{10} \leftrightarrow p_{20} \right)\bigg] \nonumber \\*
    &&{} + \frac{ih_A h_\pi}{8 \sqrt{2} F_\pi^2} P^\Phi(\tilde{k},m_K) \, \tilde{k}^\sigma \left(2 p_{10}\cdot p_{20} + \tilde{k}^2 \right).
    \label{eq:32}
\end{eqnarray}
The intermediate momenta are $k = p_a + p_\Xi$ and $\tilde{k} = p_a + p_b$, with $p_a = \left(p_+, p_0, p_{01} \right)$ and $p_b = \left(p_-, p_-, p_{02} \right)$ for decays 1-3, respectively. Note that in the momentum-swapped terms also momenta in $k$ get swapped. We use the following shorthand for the scalar, spin-1/2 and spin-3/2 propagators,
\begin{eqnarray}
    P^\Phi(p,m) &&= \frac{i}{p^2-m^2+i\epsilon}, \nonumber \\*
    P^B(p,m) &&= \frac{i\left(\slashed{p} +m \right)}{p^2-m^2+i\epsilon}, \nonumber \\*
    P^T_{\mu\nu}(p,m,\Gamma) &&= -\frac{i\left(\slashed{p} + m\right)}{p^2 - m^2 + i m \Gamma} P^{3/2}_{\mu\nu}(p) \nonumber \\*
    &&{} + \frac{2i}{3m^2}\left(\slashed{p} + m\right)\frac{p_\mu p_\nu}{p^2} \nonumber \\*
    &&{} - \frac{i}{3m} \frac{p_\mu p^\alpha \gamma_{\alpha\nu} + \gamma_{\mu\alpha} p^\alpha p_\nu}{p^2},
    \label{eq:33}
\end{eqnarray}
respectively. Propagators that do not explicitly contain a width $\Gamma$ in their argument do not hit a pole. For decay 2, we approximate $m_{\Xi^{\ast-}} \approx m_{\Xi^{\ast+}} \approx (m_{\Xi^{\ast-}} + m_{\Xi^{\ast+}})/2 = m_{\Xi^\ast}$, with an equivalent approximation for $m_\Xi$.

The Feynman matrix element for the semileptonic decay is
\begin{eqnarray}
    \mathcal{M}\left(\Omega^- \to \Xi^0 \mu^- \bar{\nu}_\mu \right) &&= \bar{u}(p_{\left(\mu\right)}) \beta'_\kappa v(p_{\left(\nu\right)}) \nonumber \\*
    &&{} \times \bar{u}(p_\Xi) \beta''^{\kappa\sigma} u_\sigma(p_\Omega),        
    \label{eq:34}
\end{eqnarray}
where
\begin{eqnarray}
    \beta'_\kappa &&= -2\sqrt{2} G_F V_{us} \gamma_\kappa P_L,\nonumber \\*
    \beta''^{\kappa\sigma} &&=  \frac{h_A}{2\sqrt{2}}g^{\kappa\sigma} + \left(c_E - c_M \gamma_5 \right)\left( \slashed{q} g^{\kappa\sigma} - \gamma^\kappa q^\sigma \right) \nonumber \\*
    &&{} - \frac{h_A q^\kappa q^\sigma }{2\sqrt{2}\left(q^2 - m_K^2\right)},
    \label{eq:35}
\end{eqnarray}
with the Fermi constant $G_F$ and $q = p_\Omega - p_\Xi$.

\section{Deriving vertices}\label{app:vertex} 
In principle, it is possible to expand the chiral Lagrangians and read off all possible interactions. However, it is much more economical to write generic vertices involving any meson or baryon \cite{Holmberg1372811,Mommers1677385}. To start, we collect the different particle fields into tuples,
\begin{eqnarray*}
    \left(B^i\right) &=& \left(\Sigma^+,\;  \Sigma^-, \;  \Sigma^0, \; p, \; \Xi^-, \; n, \; \Xi^0, \; \Lambda\right) ,\\*
    \left( \Phi^j \right) &=& \left( \pi^+, \; \pi^-, \; \pi^0, \; K^+, \; K^-, \; K^0, \; \bar{K}^0, \; \eta \right),\\*
    \left( T^n \right) &=& \big( \Sigma^{\ast +}, \; \Sigma^{\ast -}, \; \Sigma^{\ast 0}, \; \Delta^+, \; \Xi^{\ast -}, \; \Delta^0, \; \Xi^{\ast 0}, \\*
    &&{} \; \Delta^{++}, \; \Delta^-, \; \Omega^-  \big).
    \label{eq:36}
\end{eqnarray*}
Tuples for barred fields look the same.

For the $3 \times 3$ traceless matrices we use the basis $b_i$, defined via the Gell-Mann matrices as \cite{Gasser:1984gg}
\begin{eqnarray*}
    b_1 &= \frac{1}{\sqrt{2}}\left(\lambda_1 + i\lambda_2 \right),\quad b_2 &= b_1^\dagger, \quad b_3 = \lambda_3, \\*
    b_4 &= \frac{1}{\sqrt{2}}\left(\lambda_4 + i\lambda_5 \right),\quad b_5 &= b_4^\dagger, \quad    \\*
    b_6 &= \frac{1}{\sqrt{2}}\left(\lambda_6 + i\lambda_7 \right),\quad b_7 &= b_6^\dagger, \quad b_8= \lambda_8.
    \label{eq:37}
\end{eqnarray*}
From the completeness relation of the Gell-Mann matrices one may show that
\begin{equation}
  	\left(b_i\right)^a_b \left(b_i^\text{T}\right)^c_d = \left(\lambda_i\right)^a_b \left(\lambda_i\right)^c_d = 2\delta^a_d \delta^c_b - \frac{2}{3}\delta^a_b \delta^c_d.
    \label{eq:38}
\end{equation}
Note the implicit sum over $i$.

The building blocks of the totally symmetric $3 \times 3 \times 3$ tensors are the tensors $e_{abc}$, whose components are zero, except for the $abc$-element, which is one. The basis elements are then
\begin{eqnarray*}
    t_1 &&= \frac{1}{\sqrt{3}}\left( e_{113} + e_{131} + e_{311} \right), \\*
    t_2 &&= \frac{1}{\sqrt{3}}\left( e_{223} + e_{232} + e_{322} \right), \\*
    t_3 &&= \frac{1}{\sqrt{6}}\left( e_{123} + e_{132} + e_{213} + e_{231} + e_{312} + e_{321}  \right), \\*
    t_4 &&= \frac{1}{\sqrt{3}}\left( e_{112} + e_{121} + e_{211} \right), \\*
    t_5 &&= \frac{1}{\sqrt{3}}\left( e_{332} + e_{323} + e_{233} \right), \\*
    t_6 &&= \frac{1}{\sqrt{3}}\left( e_{221} + e_{212} + e_{122} \right), \\*
    t_7 &&= \frac{1}{\sqrt{3}}\left( e_{331} + e_{313} + e_{133} \right), \\*
    t_8 &&= e_{111}, \quad t_9 = e_{222}, \quad t_{10} = e_{333}.
    \label{eq:39}
\end{eqnarray*}

The fields are split from their tensor structure as
\begin{equation}
    \Phi^a_b = \Phi^j \left(b_j\right)^a_b, \quad B^a_b = \frac{1}{\sqrt{2}} B^i \left(b_i\right)^a_b, \quad T^{abc} = T^n \left(t_n\right)^{abc}.
    \label{eq:40}
\end{equation}
Note that
\begin{equation}
    \bar{B}^a_b = \frac{1}{\sqrt{2}} \bar{B}^i \left(b_i^{\text{T}}\right)^a_b.
\end{equation}

As an example, suppose we want to derive at LO the vertex between a decuplet baryon, an octet antibaryon and a meson (all in-going). The relevant part of the Lagrangian is
\[ \frac{h_A}{2\sqrt{2}} \epsilon_{ade} \bar{B}^e_c \left( u^\mu \right)^d_b T^{abc}_\mu. \]
By expanding the fields we find a vertex
\begin{equation}
    V^{T \to \Phi B}_\mu = -\frac{h_A}{2\sqrt{2} F_\pi} c_{nij} p^\Phi_\mu,
\end{equation}
where $p^\Phi_\mu$ is the meson momentum and $c_{nij}$ a `flavor factor',
\begin{equation}
    c_{nij} = \frac{1}{\sqrt{2}} \epsilon_{ade} \left(b_i\right)^e_c \left(b_j\right)^d_b \left(t_n\right)^{abc}.
\end{equation}
A straightforward evaluation of the flavor factor reveals all possible trilinear couplings.

\section{Remarks on chiral conventions}\label{app:convention}
For the SU(3) chiral Lagrangian there are two main sets of notation in use. In this work, we use notation type `A' (also used in Refs.~\cite{Gasser:1984gg,Ecker:1988te,Scherer:2002tk,Cirigliano:2011ny,Scherer:2012xha,Bijnens:2014lea,Holmberg:2018dtv,Holmberg:2019ltw}). In Refs.~\cite{Jenkins:1991es,Jenkins:1991bt,Tandean:1998ch,Borasoy:1998ku,Antipin:2007ya} notation type `B' is used. It is surprisingly messy to harmonize both types. 

In notation A, upper indices transform according to the fundamental representation and lower indices transform according to the antifundamental representation. In notation B this is flipped. Consequently, in notation B $\Phi^a_b$ refers to the $b$th row and $a$th column. Therefore, in translating notation A to notation B, upper indices become lower indices and vice versa.

From the transformation rules of the particle fields we find that fields in notation type B gain a relative phase compared to notation type A (i.e. $\eta_A = -\eta_B,\dots$). Table \ref{tab:conventions} gives a further overview. Note that (say) the matrix $\Phi_A$ contains fields of A ($\Lambda_A, \dots$), while $\Phi_B$ contains fields of B ($\Lambda_B = -\Lambda_A, \dots$).  
\begin{table}[b]
    \centering
    \caption{A comparison of two common types of notation in use for the SU(3) chiral Lagrangian. Type A is used in this work and (for example) Refs.~\cite{Gasser:1984gg,Ecker:1988te,Scherer:2002tk,Cirigliano:2011ny,Scherer:2012xha,Bijnens:2014lea,Holmberg:2018dtv,Holmberg:2019ltw} while type B is used in Refs.~\cite{Jenkins:1991es,Jenkins:1991bt,Tandean:1998ch,Borasoy:1998ku,Antipin:2007ya}.}
    \label{tab:conventions}
    \begin{ruledtabular}
        \begin{tabular}{ c c | c c }
            Type A & Type B & Type A & Type B\\ \hline
            $B$                                                & $B$           		& $D^\mu$                                            & $\mathcal{D}^\mu$     	\\
            $\Phi$                                             & $2 \pi$            & $F_\pi$                                            & $f$    \\
            $U$                                                & $\Sigma^\dagger$   & $m_{\left(8\right)}$                               & $m_B$      \\
            $u$                                                & $\xi^\dagger$      & $m_{\left(10\right)}$                              & $m_T$    \\
            $L/R$                                              & $L/R$              & $D$                                                & $D$                        \\
            $h_V$                                                & $U$                & $F$                                                & $F$        \\
            $\Gamma_\mu \big\rvert_{l_\nu, r_\nu = 0}$         & $V_\mu$            & $h_A$                                              & $\sqrt{2} \mathcal{C}$     \\
            $ u_\mu  \big\rvert_{l_\nu, r_\nu = 0} $           & $2 A_\mu$          & $H_A$                                              & $-\mathcal{H}$     \\                                                        
        \end{tabular}
    \end{ruledtabular}
\end{table}

\bibliographystyle{apsrev4-2}
\bibliography{lit}

\begin{thebibliography}{60}%
\makeatletter
\providecommand \@ifxundefined [1]{%
 \@ifx{#1\undefined}
}%
\providecommand \@ifnum [1]{%
 \ifnum #1\expandafter \@firstoftwo
 \else \expandafter \@secondoftwo
 \fi
}%
\providecommand \@ifx [1]{%
 \ifx #1\expandafter \@firstoftwo
 \else \expandafter \@secondoftwo
 \fi
}%
\providecommand \natexlab [1]{#1}%
\providecommand \enquote  [1]{``#1''}%
\providecommand \bibnamefont  [1]{#1}%
\providecommand \bibfnamefont [1]{#1}%
\providecommand \citenamefont [1]{#1}%
\providecommand \href@noop [0]{\@secondoftwo}%
\providecommand \href [0]{\begingroup \@sanitize@url \@href}%
\providecommand \@href[1]{\@@startlink{#1}\@@href}%
\providecommand \@@href[1]{\endgroup#1\@@endlink}%
\providecommand \@sanitize@url [0]{\catcode `\\12\catcode `\$12\catcode
  `\&12\catcode `\#12\catcode `\^12\catcode `\_12\catcode `\%12\relax}%
\providecommand \@@startlink[1]{}%
\providecommand \@@endlink[0]{}%
\providecommand \url  [0]{\begingroup\@sanitize@url \@url }%
\providecommand \@url [1]{\endgroup\@href {#1}{\urlprefix }}%
\providecommand \urlprefix  [0]{URL }%
\providecommand \Eprint [0]{\href }%
\providecommand \doibase [0]{https://doi.org/}%
\providecommand \selectlanguage [0]{\@gobble}%
\providecommand \bibinfo  [0]{\@secondoftwo}%
\providecommand \bibfield  [0]{\@secondoftwo}%
\providecommand \translation [1]{[#1]}%
\providecommand \BibitemOpen [0]{}%
\providecommand \bibitemStop [0]{}%
\providecommand \bibitemNoStop [0]{.\EOS\space}%
\providecommand \EOS [0]{\spacefactor3000\relax}%
\providecommand \BibitemShut  [1]{\csname bibitem#1\endcsname}%
\let\auto@bib@innerbib\@empty
\bibitem [{\citenamefont {Bigi}\ and\ \citenamefont {Sanda}(2016)}]{bigi-CP}%
  \BibitemOpen
  \bibfield  {author} {\bibinfo {author} {\bibfnamefont {I.}~\bibnamefont
  {Bigi}}\ and\ \bibinfo {author} {\bibfnamefont {A.}~\bibnamefont {Sanda}},\
  }\href@noop {} {\emph {\bibinfo {title} {CP Violation}}},\ \bibinfo {edition}
  {2nd}\ ed.,\ Cambridge Monographs on Particle Physics, Nuclear Physics and
  Cosmology\ (\bibinfo  {publisher} {Cambridge Univ. Press},\ \bibinfo {year}
  {2016})\BibitemShut {NoStop}%
\bibitem [{\citenamefont {Donoghue}\ \emph {et~al.}(2014)\citenamefont
  {Donoghue}, \citenamefont {Golowich},\ and\ \citenamefont
  {Holstein}}]{Donoghue:1992dd}%
  \BibitemOpen
  \bibfield  {author} {\bibinfo {author} {\bibfnamefont {J.~F.}\ \bibnamefont
  {Donoghue}}, \bibinfo {author} {\bibfnamefont {E.}~\bibnamefont {Golowich}},\
  and\ \bibinfo {author} {\bibfnamefont {B.~R.}\ \bibnamefont {Holstein}},\
  }\href {https://doi.org/10.1017/CBO9780511524370} {\emph {\bibinfo {title}
  {{Dynamics of the standard model}}}},\ Cambridge Monographs on Particle
  Physics, Nuclear Physics and Cosmology\ (\bibinfo  {publisher} {Cambridge
  Univ. Press},\ \bibinfo {year} {2014})\BibitemShut {NoStop}%
\bibitem [{\citenamefont {Cirigliano}\ \emph {et~al.}(2012)\citenamefont
  {Cirigliano}, \citenamefont {Ecker}, \citenamefont {Neufeld}, \citenamefont
  {Pich},\ and\ \citenamefont {Portoles}}]{Cirigliano:2011ny}%
  \BibitemOpen
  \bibfield  {author} {\bibinfo {author} {\bibfnamefont {V.}~\bibnamefont
  {Cirigliano}}, \bibinfo {author} {\bibfnamefont {G.}~\bibnamefont {Ecker}},
  \bibinfo {author} {\bibfnamefont {H.}~\bibnamefont {Neufeld}}, \bibinfo
  {author} {\bibfnamefont {A.}~\bibnamefont {Pich}},\ and\ \bibinfo {author}
  {\bibfnamefont {J.}~\bibnamefont {Portoles}},\ }\href
  {https://doi.org/10.1103/RevModPhys.84.399} {\bibfield  {journal} {\bibinfo
  {journal} {Rev. Mod. Phys.}\ }\textbf {\bibinfo {volume} {84}},\ \bibinfo
  {pages} {399} (\bibinfo {year} {2012})},\ \Eprint
  {https://arxiv.org/abs/1107.6001} {arXiv:1107.6001 [hep-ph]} \BibitemShut
  {NoStop}%
\bibitem [{\citenamefont {Salone}\ \emph {et~al.}(2022)\citenamefont {Salone},
  \citenamefont {Adlarson}, \citenamefont {Batozskaya}, \citenamefont {Kupsc},
  \citenamefont {Leupold},\ and\ \citenamefont {Tandean}}]{Salone:2022lpt}%
  \BibitemOpen
  \bibfield  {author} {\bibinfo {author} {\bibfnamefont {N.}~\bibnamefont
  {Salone}}, \bibinfo {author} {\bibfnamefont {P.}~\bibnamefont {Adlarson}},
  \bibinfo {author} {\bibfnamefont {V.}~\bibnamefont {Batozskaya}}, \bibinfo
  {author} {\bibfnamefont {A.}~\bibnamefont {Kupsc}}, \bibinfo {author}
  {\bibfnamefont {S.}~\bibnamefont {Leupold}},\ and\ \bibinfo {author}
  {\bibfnamefont {J.}~\bibnamefont {Tandean}},\ }\href
  {https://doi.org/10.1103/PhysRevD.105.116022} {\bibfield  {journal} {\bibinfo
   {journal} {Phys. Rev. D}\ }\textbf {\bibinfo {volume} {105}},\ \bibinfo
  {pages} {116022} (\bibinfo {year} {2022})},\ \Eprint
  {https://arxiv.org/abs/2203.03035} {arXiv:2203.03035 [hep-ph]} \BibitemShut
  {NoStop}%
\bibitem [{\citenamefont {Workman}\ \emph {et~al.}(2022)\citenamefont {Workman}
  \emph {et~al.}}]{pdg}%
  \BibitemOpen
  \bibfield  {author} {\bibinfo {author} {\bibfnamefont {R.~L.}\ \bibnamefont
  {Workman}} \emph {et~al.} (\bibinfo {collaboration} {Particle Data Group}),\
  }\href {https://doi.org/10.1093/ptep/ptac097} {\bibfield  {journal} {\bibinfo
   {journal} {PTEP}\ }\textbf {\bibinfo {volume} {2022}},\ \bibinfo {pages}
  {083C01} (\bibinfo {year} {2022})}\BibitemShut {NoStop}%
\bibitem [{\citenamefont {Goswami}\ and\ \citenamefont
  {Schechter}(1970)}]{Goswami:1970wqc}%
  \BibitemOpen
  \bibfield  {author} {\bibinfo {author} {\bibfnamefont {D.~N.}\ \bibnamefont
  {Goswami}}\ and\ \bibinfo {author} {\bibfnamefont {J.}~\bibnamefont
  {Schechter}},\ }\href {https://doi.org/10.1103/PhysRevD.1.290} {\bibfield
  {journal} {\bibinfo  {journal} {Phys. Rev. D}\ }\textbf {\bibinfo {volume}
  {1}},\ \bibinfo {pages} {290} (\bibinfo {year} {1970})},\ \bibinfo {note}
  {[Erratum: Phys.Rev.D 4, 3526 (1971)]}\BibitemShut {NoStop}%
\bibitem [{\citenamefont {Carone}\ and\ \citenamefont
  {Georgi}(1992)}]{Carone:1991ni}%
  \BibitemOpen
  \bibfield  {author} {\bibinfo {author} {\bibfnamefont {C.}~\bibnamefont
  {Carone}}\ and\ \bibinfo {author} {\bibfnamefont {H.}~\bibnamefont
  {Georgi}},\ }\href {https://doi.org/10.1016/0550-3213(92)90031-6} {\bibfield
  {journal} {\bibinfo  {journal} {Nucl. Phys. B}\ }\textbf {\bibinfo {volume}
  {375}},\ \bibinfo {pages} {243} (\bibinfo {year} {1992})}\BibitemShut
  {NoStop}%
\bibitem [{\citenamefont {Tandean}\ and\ \citenamefont
  {Valencia}(1999)}]{Tandean:1998ch}%
  \BibitemOpen
  \bibfield  {author} {\bibinfo {author} {\bibfnamefont {J.}~\bibnamefont
  {Tandean}}\ and\ \bibinfo {author} {\bibfnamefont {G.}~\bibnamefont
  {Valencia}},\ }\href {https://doi.org/10.1016/S0370-2693(99)00277-4}
  {\bibfield  {journal} {\bibinfo  {journal} {Phys. Lett. B}\ }\textbf
  {\bibinfo {volume} {452}},\ \bibinfo {pages} {395} (\bibinfo {year}
  {1999})},\ \Eprint {https://arxiv.org/abs/hep-ph/9810201}
  {arXiv:hep-ph/9810201} \BibitemShut {NoStop}%
\bibitem [{\citenamefont {Egolf}\ \emph {et~al.}(1999)\citenamefont {Egolf},
  \citenamefont {Melnikov},\ and\ \citenamefont {Springer}}]{Egolf:1998vj}%
  \BibitemOpen
  \bibfield  {author} {\bibinfo {author} {\bibfnamefont {D.~A.}\ \bibnamefont
  {Egolf}}, \bibinfo {author} {\bibfnamefont {I.~V.}\ \bibnamefont
  {Melnikov}},\ and\ \bibinfo {author} {\bibfnamefont {R.~P.}\ \bibnamefont
  {Springer}},\ }\href {https://doi.org/10.1016/S0370-2693(99)00234-8}
  {\bibfield  {journal} {\bibinfo  {journal} {Phys. Lett. B}\ }\textbf
  {\bibinfo {volume} {451}},\ \bibinfo {pages} {267} (\bibinfo {year}
  {1999})},\ \Eprint {https://arxiv.org/abs/hep-ph/9809228}
  {arXiv:hep-ph/9809228} \BibitemShut {NoStop}%
\bibitem [{\citenamefont {Antipin}\ \emph {et~al.}(2007)\citenamefont
  {Antipin}, \citenamefont {Tandean},\ and\ \citenamefont
  {Valencia}}]{Antipin:2007ya}%
  \BibitemOpen
  \bibfield  {author} {\bibinfo {author} {\bibfnamefont {O.}~\bibnamefont
  {Antipin}}, \bibinfo {author} {\bibfnamefont {J.}~\bibnamefont {Tandean}},\
  and\ \bibinfo {author} {\bibfnamefont {G.}~\bibnamefont {Valencia}},\ }\href
  {https://doi.org/10.1103/PhysRevD.76.094024} {\bibfield  {journal} {\bibinfo
  {journal} {Phys. Rev. D}\ }\textbf {\bibinfo {volume} {76}},\ \bibinfo
  {pages} {094024} (\bibinfo {year} {2007})},\ \Eprint
  {https://arxiv.org/abs/0705.3279} {arXiv:0705.3279 [hep-ph]} \BibitemShut
  {NoStop}%
\bibitem [{\citenamefont {Bourquin}\ \emph {et~al.}(1984)\citenamefont
  {Bourquin} \emph {et~al.}}]{Bourquin19841}%
  \BibitemOpen
  \bibfield  {author} {\bibinfo {author} {\bibfnamefont {M.}~\bibnamefont
  {Bourquin}} \emph {et~al.} (\bibinfo {collaboration}
  {Bristol-Geneva-Heidelberg-Orsay-Rutherford-Strasbourg}),\ }\href
  {https://doi.org/10.1016/0550-3213(84)90195-0} {\bibfield  {journal}
  {\bibinfo  {journal} {Nucl. Phys. B}\ }\textbf {\bibinfo {volume} {241}},\
  \bibinfo {pages} {1} (\bibinfo {year} {1984})}\BibitemShut {NoStop}%
\bibitem [{\citenamefont {Kamaev}\ \emph {et~al.}(2010)\citenamefont {Kamaev}
  \emph {et~al.}}]{HyperCP:2010ego}%
  \BibitemOpen
  \bibfield  {author} {\bibinfo {author} {\bibfnamefont {O.}~\bibnamefont
  {Kamaev}} \emph {et~al.} (\bibinfo {collaboration} {HyperCP}),\ }\href
  {https://doi.org/10.1016/j.physletb.2010.08.037} {\bibfield  {journal}
  {\bibinfo  {journal} {Phys. Lett. B}\ }\textbf {\bibinfo {volume} {693}},\
  \bibinfo {pages} {236} (\bibinfo {year} {2010})},\ \Eprint
  {https://arxiv.org/abs/1008.4405} {arXiv:1008.4405 [hep-ex]} \BibitemShut
  {NoStop}%
\bibitem [{\citenamefont {Asner}\ \emph {et~al.}(2009)\citenamefont {Asner}
  \emph {et~al.}}]{Asner2008}%
  \BibitemOpen
  \bibfield  {author} {\bibinfo {author} {\bibfnamefont {D.~M.}\ \bibnamefont
  {Asner}} \emph {et~al.},\ }\href@noop {} {\bibfield  {journal} {\bibinfo
  {journal} {Int. J. Mod. Phys. A}\ }\textbf {\bibinfo {volume} {24}},\
  \bibinfo {pages} {S1} (\bibinfo {year} {2009})},\ \Eprint
  {https://arxiv.org/abs/0809.1869} {arXiv:0809.1869 [hep-ex]} \BibitemShut
  {NoStop}%
\bibitem [{\citenamefont {Ablikim}\ \emph {et~al.}(2020)\citenamefont {Ablikim}
  \emph {et~al.}}]{BESIII2020}%
  \BibitemOpen
  \bibfield  {author} {\bibinfo {author} {\bibfnamefont {M.}~\bibnamefont
  {Ablikim}} \emph {et~al.} (\bibinfo {collaboration} {BESIII}),\ }\href
  {https://doi.org/10.1088/1674-1137/44/4/040001} {\bibfield  {journal}
  {\bibinfo  {journal} {Chin. Phys. C}\ }\textbf {\bibinfo {volume} {44}},\
  \bibinfo {pages} {040001} (\bibinfo {year} {2020})},\ \Eprint
  {https://arxiv.org/abs/1912.05983} {arXiv:1912.05983 [hep-ex]} \BibitemShut
  {NoStop}%
\bibitem [{\citenamefont {Ablikim}\ \emph {et~al.}(2021)\citenamefont {Ablikim}
  \emph {et~al.}}]{BESIII:2020lkm}%
  \BibitemOpen
  \bibfield  {author} {\bibinfo {author} {\bibfnamefont {M.}~\bibnamefont
  {Ablikim}} \emph {et~al.} (\bibinfo {collaboration} {BESIII}),\ }\href
  {https://doi.org/10.1103/PhysRevLett.126.092002} {\bibfield  {journal}
  {\bibinfo  {journal} {Phys. Rev. Lett.}\ }\textbf {\bibinfo {volume} {126}},\
  \bibinfo {pages} {092002} (\bibinfo {year} {2021})},\ \Eprint
  {https://arxiv.org/abs/2007.03679} {arXiv:2007.03679 [hep-ex]} \BibitemShut
  {NoStop}%
\bibitem [{\citenamefont {Belyaev}\ \emph {et~al.}(2021)\citenamefont
  {Belyaev}, \citenamefont {Carboni}, \citenamefont {Harnew}, \citenamefont
  {Matteuzzi},\ and\ \citenamefont {Teubert}}]{Belyaev2021}%
  \BibitemOpen
  \bibfield  {author} {\bibinfo {author} {\bibfnamefont {I.}~\bibnamefont
  {Belyaev}}, \bibinfo {author} {\bibfnamefont {G.}~\bibnamefont {Carboni}},
  \bibinfo {author} {\bibfnamefont {N.}~\bibnamefont {Harnew}}, \bibinfo
  {author} {\bibfnamefont {C.}~\bibnamefont {Matteuzzi}},\ and\ \bibinfo
  {author} {\bibfnamefont {F.}~\bibnamefont {Teubert}},\ }\href
  {https://doi.org/10.1140/epjh/s13129-021-00002-z} {\bibfield  {journal}
  {\bibinfo  {journal} {Eur. Phys. J. H}\ }\textbf {\bibinfo {volume} {46}},\
  \bibinfo {pages} {3} (\bibinfo {year} {2021})},\ \Eprint
  {https://arxiv.org/abs/2101.05331} {arXiv:2101.05331 [physics.hist-ph]}
  \BibitemShut {NoStop}%
\bibitem [{\citenamefont {Abe}\ \emph {et~al.}(2010)\citenamefont {Abe} \emph
  {et~al.}}]{Belle-II2010}%
  \BibitemOpen
  \bibfield  {author} {\bibinfo {author} {\bibfnamefont {T.}~\bibnamefont
  {Abe}} \emph {et~al.} (\bibinfo {collaboration} {Belle-II}),\ }\href@noop {}
  {\  (\bibinfo {year} {2010})},\ \bibinfo {note} {{Belle II Technical Design
  Report}},\ \Eprint {https://arxiv.org/abs/1011.0352} {arXiv:1011.0352
  [physics.ins-det]} \BibitemShut {NoStop}%
\bibitem [{\citenamefont {Altmannshofer}\ \emph {et~al.}(2019)\citenamefont
  {Altmannshofer} \emph {et~al.}}]{Belle-II:2018jsg}%
  \BibitemOpen
  \bibfield  {author} {\bibinfo {author} {\bibfnamefont {W.}~\bibnamefont
  {Altmannshofer}} \emph {et~al.} (\bibinfo {collaboration} {Belle-II}),\
  }\href {https://doi.org/10.1093/ptep/ptz106} {\bibfield  {journal} {\bibinfo
  {journal} {PTEP}\ }\textbf {\bibinfo {volume} {2019}},\ \bibinfo {pages}
  {123C01} (\bibinfo {year} {2019})},\ \bibinfo {note} {[Erratum: PTEP 2020,
  029201 (2020)]},\ \Eprint {https://arxiv.org/abs/1808.10567}
  {arXiv:1808.10567 [hep-ex]} \BibitemShut {NoStop}%
\bibitem [{\citenamefont {Lutz}\ \emph {et~al.}(2009)\citenamefont {Lutz} \emph
  {et~al.}}]{PANDA2009}%
  \BibitemOpen
  \bibfield  {author} {\bibinfo {author} {\bibfnamefont {M.~F.~M.}\
  \bibnamefont {Lutz}} \emph {et~al.} (\bibinfo {collaboration} {PANDA}),\
  }\href@noop {} {\  (\bibinfo {year} {2009})},\ \bibinfo {note} {{Physics
  Performance Report for PANDA: Strong Interaction Studies with Antiprotons}},\
  \Eprint {https://arxiv.org/abs/0903.3905} {arXiv:0903.3905 [hep-ex]}
  \BibitemShut {NoStop}%
\bibitem [{\citenamefont {Barucca}\ \emph {et~al.}(2021)\citenamefont {Barucca}
  \emph {et~al.}}]{PANDA2021}%
  \BibitemOpen
  \bibfield  {author} {\bibinfo {author} {\bibfnamefont {G.}~\bibnamefont
  {Barucca}} \emph {et~al.} (\bibinfo {collaboration} {PANDA}),\ }\href
  {https://doi.org/10.1140/epja/s10050-021-00475-y} {\bibfield  {journal}
  {\bibinfo  {journal} {Eur. Phys. J. A}\ }\textbf {\bibinfo {volume} {57}},\
  \bibinfo {pages} {184} (\bibinfo {year} {2021})},\ \Eprint
  {https://arxiv.org/abs/2101.11877} {arXiv:2101.11877 [hep-ex]} \BibitemShut
  {NoStop}%
\bibitem [{\citenamefont {Holmberg}\ and\ \citenamefont
  {Leupold}(2018)}]{Holmberg:2018dtv}%
  \BibitemOpen
  \bibfield  {author} {\bibinfo {author} {\bibfnamefont {M.}~\bibnamefont
  {Holmberg}}\ and\ \bibinfo {author} {\bibfnamefont {S.}~\bibnamefont
  {Leupold}},\ }\href {https://doi.org/10.1140/epja/i2018-12533-3} {\bibfield
  {journal} {\bibinfo  {journal} {Eur. Phys. J.}\ }\textbf {\bibinfo {volume}
  {A54}},\ \bibinfo {pages} {103} (\bibinfo {year} {2018})},\ \Eprint
  {https://arxiv.org/abs/1802.05168} {arXiv:1802.05168 [hep-ph]} \BibitemShut
  {NoStop}%
\bibitem [{\citenamefont {Holmberg}\ and\ \citenamefont
  {Leupold}(2019)}]{Holmberg:2019ltw}%
  \BibitemOpen
  \bibfield  {author} {\bibinfo {author} {\bibfnamefont {M.}~\bibnamefont
  {Holmberg}}\ and\ \bibinfo {author} {\bibfnamefont {S.}~\bibnamefont
  {Leupold}},\ }\href {https://doi.org/10.1103/PhysRevD.100.114001} {\bibfield
  {journal} {\bibinfo  {journal} {Phys. Rev. D}\ }\textbf {\bibinfo {volume}
  {100}},\ \bibinfo {pages} {114001} (\bibinfo {year} {2019})},\ \Eprint
  {https://arxiv.org/abs/1909.13562} {arXiv:1909.13562 [hep-ph]} \BibitemShut
  {NoStop}%
\bibitem [{\citenamefont {Alvarez-Ruso}\ \emph {et~al.}(2018)\citenamefont
  {Alvarez-Ruso} \emph {et~al.}}]{NuSTEC2017}%
  \BibitemOpen
  \bibfield  {author} {\bibinfo {author} {\bibfnamefont {L.}~\bibnamefont
  {Alvarez-Ruso}} \emph {et~al.} (\bibinfo {collaboration} {NuSTEC}),\ }\href
  {https://doi.org/10.1016/j.ppnp.2018.01.006} {\bibfield  {journal} {\bibinfo
  {journal} {Prog. Part. Nucl. Phys.}\ }\textbf {\bibinfo {volume} {100}},\
  \bibinfo {pages} {1} (\bibinfo {year} {2018})},\ \Eprint
  {https://arxiv.org/abs/1706.03621} {arXiv:1706.03621 [hep-ph]} \BibitemShut
  {NoStop}%
\bibitem [{\citenamefont {Procura}(2008)}]{Procura2008}%
  \BibitemOpen
  \bibfield  {author} {\bibinfo {author} {\bibfnamefont {M.}~\bibnamefont
  {Procura}},\ }\href {https://doi.org/10.1103/PhysRevD.78.094021} {\bibfield
  {journal} {\bibinfo  {journal} {Phys. Rev. D}\ }\textbf {\bibinfo {volume}
  {78}},\ \bibinfo {pages} {094021} (\bibinfo {year} {2008})},\ \Eprint
  {https://arxiv.org/abs/0803.4291} {arXiv:0803.4291 [hep-ph]} \BibitemShut
  {NoStop}%
\bibitem [{\citenamefont {Geng}\ \emph
  {et~al.}(2008{\natexlab{a}})\citenamefont {Geng}, \citenamefont
  {Martin~Camalich}, \citenamefont {Alvarez-Ruso},\ and\ \citenamefont
  {Vicente~Vacas}}]{Geng2008}%
  \BibitemOpen
  \bibfield  {author} {\bibinfo {author} {\bibfnamefont {L.~S.}\ \bibnamefont
  {Geng}}, \bibinfo {author} {\bibfnamefont {J.}~\bibnamefont
  {Martin~Camalich}}, \bibinfo {author} {\bibfnamefont {L.}~\bibnamefont
  {Alvarez-Ruso}},\ and\ \bibinfo {author} {\bibfnamefont {M.~J.}\ \bibnamefont
  {Vicente~Vacas}},\ }\href {https://doi.org/10.1103/PhysRevD.78.014011}
  {\bibfield  {journal} {\bibinfo  {journal} {Phys. Rev. D}\ }\textbf {\bibinfo
  {volume} {78}},\ \bibinfo {pages} {014011} (\bibinfo {year}
  {2008}{\natexlab{a}})},\ \Eprint {https://arxiv.org/abs/0801.4495}
  {arXiv:0801.4495 [hep-ph]} \BibitemShut {NoStop}%
\bibitem [{\citenamefont {Mosel}(2015)}]{Mosel2015}%
  \BibitemOpen
  \bibfield  {author} {\bibinfo {author} {\bibfnamefont {U.}~\bibnamefont
  {Mosel}},\ }\href {https://doi.org/10.1103/PhysRevC.91.065501} {\bibfield
  {journal} {\bibinfo  {journal} {Phys. Rev. C}\ }\textbf {\bibinfo {volume}
  {91}},\ \bibinfo {pages} {065501} (\bibinfo {year} {2015})},\ \Eprint
  {https://arxiv.org/abs/1502.08032} {arXiv:1502.08032 [nucl-th]} \BibitemShut
  {NoStop}%
\bibitem [{\citenamefont {\"Unal}\ \emph {et~al.}(2019)\citenamefont {\"Unal},
  \citenamefont {K\"u\c{c}\"ukarslan},\ and\ \citenamefont
  {Scherer}}]{Unal2018}%
  \BibitemOpen
  \bibfield  {author} {\bibinfo {author} {\bibfnamefont {Y.}~\bibnamefont
  {\"Unal}}, \bibinfo {author} {\bibfnamefont {A.}~\bibnamefont
  {K\"u\c{c}\"ukarslan}},\ and\ \bibinfo {author} {\bibfnamefont
  {S.}~\bibnamefont {Scherer}},\ }\href
  {https://doi.org/10.1103/PhysRevD.99.014012} {\bibfield  {journal} {\bibinfo
  {journal} {Phys. Rev. D}\ }\textbf {\bibinfo {volume} {99}},\ \bibinfo
  {pages} {014012} (\bibinfo {year} {2019})},\ \Eprint
  {https://arxiv.org/abs/1808.03046} {arXiv:1808.03046 [hep-ph]} \BibitemShut
  {NoStop}%
\bibitem [{\citenamefont {Antonelli}\ \emph {et~al.}(1996)\citenamefont
  {Antonelli}, \citenamefont {Bertolini}, \citenamefont {Eeg}, \citenamefont
  {Fabbrichesi},\ and\ \citenamefont {Lashin}}]{Antonelli:1995nv}%
  \BibitemOpen
  \bibfield  {author} {\bibinfo {author} {\bibfnamefont {V.}~\bibnamefont
  {Antonelli}}, \bibinfo {author} {\bibfnamefont {S.}~\bibnamefont
  {Bertolini}}, \bibinfo {author} {\bibfnamefont {J.~O.}\ \bibnamefont {Eeg}},
  \bibinfo {author} {\bibfnamefont {M.}~\bibnamefont {Fabbrichesi}},\ and\
  \bibinfo {author} {\bibfnamefont {E.~I.}\ \bibnamefont {Lashin}},\ }\href
  {https://doi.org/10.1016/0550-3213(96)00144-7} {\bibfield  {journal}
  {\bibinfo  {journal} {Nucl. Phys. B}\ }\textbf {\bibinfo {volume} {469}},\
  \bibinfo {pages} {143} (\bibinfo {year} {1996})},\ \Eprint
  {https://arxiv.org/abs/hep-ph/9511255} {arXiv:hep-ph/9511255} \BibitemShut
  {NoStop}%
\bibitem [{\citenamefont {Gasser}\ and\ \citenamefont
  {Leutwyler}(1985)}]{Gasser:1984gg}%
  \BibitemOpen
  \bibfield  {author} {\bibinfo {author} {\bibfnamefont {J.}~\bibnamefont
  {Gasser}}\ and\ \bibinfo {author} {\bibfnamefont {H.}~\bibnamefont
  {Leutwyler}},\ }\href {https://doi.org/10.1016/0550-3213(85)90492-4}
  {\bibfield  {journal} {\bibinfo  {journal} {Nucl. Phys.}\ }\textbf {\bibinfo
  {volume} {B250}},\ \bibinfo {pages} {465} (\bibinfo {year}
  {1985})}\BibitemShut {NoStop}%
\bibitem [{\citenamefont {Ecker}\ \emph {et~al.}(1989)\citenamefont {Ecker},
  \citenamefont {Gasser}, \citenamefont {Pich},\ and\ \citenamefont
  {de~Rafael}}]{Ecker:1988te}%
  \BibitemOpen
  \bibfield  {author} {\bibinfo {author} {\bibfnamefont {G.}~\bibnamefont
  {Ecker}}, \bibinfo {author} {\bibfnamefont {J.}~\bibnamefont {Gasser}},
  \bibinfo {author} {\bibfnamefont {A.}~\bibnamefont {Pich}},\ and\ \bibinfo
  {author} {\bibfnamefont {E.}~\bibnamefont {de~Rafael}},\ }\href
  {https://doi.org/10.1016/0550-3213(89)90346-5} {\bibfield  {journal}
  {\bibinfo  {journal} {Nucl. Phys.}\ }\textbf {\bibinfo {volume} {B321}},\
  \bibinfo {pages} {311} (\bibinfo {year} {1989})}\BibitemShut {NoStop}%
\bibitem [{\citenamefont {Donoghue}\ \emph {et~al.}(1989)\citenamefont
  {Donoghue}, \citenamefont {Ramirez},\ and\ \citenamefont
  {Valencia}}]{Donoghue:1988ed}%
  \BibitemOpen
  \bibfield  {author} {\bibinfo {author} {\bibfnamefont {J.~F.}\ \bibnamefont
  {Donoghue}}, \bibinfo {author} {\bibfnamefont {C.}~\bibnamefont {Ramirez}},\
  and\ \bibinfo {author} {\bibfnamefont {G.}~\bibnamefont {Valencia}},\ }\href
  {https://doi.org/10.1103/PhysRevD.39.1947} {\bibfield  {journal} {\bibinfo
  {journal} {Phys. Rev. D}\ }\textbf {\bibinfo {volume} {39}},\ \bibinfo
  {pages} {1947} (\bibinfo {year} {1989})}\BibitemShut {NoStop}%
\bibitem [{\citenamefont {Bijnens}\ and\ \citenamefont
  {Ecker}(2014)}]{Bijnens:2014lea}%
  \BibitemOpen
  \bibfield  {author} {\bibinfo {author} {\bibfnamefont {J.}~\bibnamefont
  {Bijnens}}\ and\ \bibinfo {author} {\bibfnamefont {G.}~\bibnamefont
  {Ecker}},\ }\href {https://doi.org/10.1146/annurev-nucl-102313-025528}
  {\bibfield  {journal} {\bibinfo  {journal} {Ann. Rev. Nucl. Part. Sci.}\
  }\textbf {\bibinfo {volume} {64}},\ \bibinfo {pages} {149} (\bibinfo {year}
  {2014})},\ \Eprint {https://arxiv.org/abs/1405.6488} {arXiv:1405.6488
  [hep-ph]} \BibitemShut {NoStop}%
\bibitem [{\citenamefont {Ellis}\ \emph {et~al.}(1977)\citenamefont {Ellis},
  \citenamefont {Gaillard}, \citenamefont {Nanopoulos},\ and\ \citenamefont
  {Rudaz}}]{Ellis:1977uk}%
  \BibitemOpen
  \bibfield  {author} {\bibinfo {author} {\bibfnamefont {J.~R.}\ \bibnamefont
  {Ellis}}, \bibinfo {author} {\bibfnamefont {M.~K.}\ \bibnamefont {Gaillard}},
  \bibinfo {author} {\bibfnamefont {D.~V.}\ \bibnamefont {Nanopoulos}},\ and\
  \bibinfo {author} {\bibfnamefont {S.}~\bibnamefont {Rudaz}},\ }\href
  {https://doi.org/10.1016/0550-3213(77)90374-1} {\bibfield  {journal}
  {\bibinfo  {journal} {Nucl. Phys. B}\ }\textbf {\bibinfo {volume} {131}},\
  \bibinfo {pages} {285} (\bibinfo {year} {1977})},\ \bibinfo {note} {[Erratum:
  Nucl.Phys.B 132, 541 (1978)]}\BibitemShut {NoStop}%
\bibitem [{\citenamefont {Jenkins}(1992)}]{Jenkins:1991bt}%
  \BibitemOpen
  \bibfield  {author} {\bibinfo {author} {\bibfnamefont {E.~E.}\ \bibnamefont
  {Jenkins}},\ }\href {https://doi.org/10.1016/0550-3213(92)90111-N} {\bibfield
   {journal} {\bibinfo  {journal} {Nucl. Phys. B}\ }\textbf {\bibinfo {volume}
  {375}},\ \bibinfo {pages} {561} (\bibinfo {year} {1992})}\BibitemShut
  {NoStop}%
\bibitem [{\citenamefont {Geng}\ \emph
  {et~al.}(2008{\natexlab{b}})\citenamefont {Geng}, \citenamefont
  {Martin~Camalich}, \citenamefont {Alvarez-Ruso},\ and\ \citenamefont
  {Vicente~Vacas}}]{Geng:2008mf}%
  \BibitemOpen
  \bibfield  {author} {\bibinfo {author} {\bibfnamefont {L.~S.}\ \bibnamefont
  {Geng}}, \bibinfo {author} {\bibfnamefont {J.}~\bibnamefont
  {Martin~Camalich}}, \bibinfo {author} {\bibfnamefont {L.}~\bibnamefont
  {Alvarez-Ruso}},\ and\ \bibinfo {author} {\bibfnamefont {M.~J.}\ \bibnamefont
  {Vicente~Vacas}},\ }\href {https://doi.org/10.1103/PhysRevLett.101.222002}
  {\bibfield  {journal} {\bibinfo  {journal} {Phys. Rev. Lett.}\ }\textbf
  {\bibinfo {volume} {101}},\ \bibinfo {pages} {222002} (\bibinfo {year}
  {2008}{\natexlab{b}})},\ \Eprint {https://arxiv.org/abs/0805.1419}
  {arXiv:0805.1419 [hep-ph]} \BibitemShut {NoStop}%
\bibitem [{\citenamefont {Lutz}\ and\ \citenamefont
  {Kolomeitsev}(2002)}]{Lutz:2001yb}%
  \BibitemOpen
  \bibfield  {author} {\bibinfo {author} {\bibfnamefont {M.~F.~M.}\
  \bibnamefont {Lutz}}\ and\ \bibinfo {author} {\bibfnamefont {E.~E.}\
  \bibnamefont {Kolomeitsev}},\ }\href
  {https://doi.org/10.1016/S0375-9474(01)01312-4} {\bibfield  {journal}
  {\bibinfo  {journal} {Nucl. Phys.}\ }\textbf {\bibinfo {volume} {A700}},\
  \bibinfo {pages} {193} (\bibinfo {year} {2002})},\ \Eprint
  {https://arxiv.org/abs/nucl-th/0105042} {arXiv:nucl-th/0105042 [nucl-th]}
  \BibitemShut {NoStop}%
\bibitem [{\citenamefont {Semke}\ and\ \citenamefont
  {Lutz}(2006)}]{Semke:2005sn}%
  \BibitemOpen
  \bibfield  {author} {\bibinfo {author} {\bibfnamefont {A.}~\bibnamefont
  {Semke}}\ and\ \bibinfo {author} {\bibfnamefont {M.~F.~M.}\ \bibnamefont
  {Lutz}},\ }\href {https://doi.org/10.1016/j.nuclphysa.2006.07.043} {\bibfield
   {journal} {\bibinfo  {journal} {Nucl. Phys.}\ }\textbf {\bibinfo {volume}
  {A778}},\ \bibinfo {pages} {153} (\bibinfo {year} {2006})},\ \Eprint
  {https://arxiv.org/abs/nucl-th/0511061} {arXiv:nucl-th/0511061 [nucl-th]}
  \BibitemShut {NoStop}%
\bibitem [{\citenamefont {Pascalutsa}\ \emph {et~al.}(2007)\citenamefont
  {Pascalutsa}, \citenamefont {Vanderhaeghen},\ and\ \citenamefont
  {Yang}}]{Pascalutsa:2006up}%
  \BibitemOpen
  \bibfield  {author} {\bibinfo {author} {\bibfnamefont {V.}~\bibnamefont
  {Pascalutsa}}, \bibinfo {author} {\bibfnamefont {M.}~\bibnamefont
  {Vanderhaeghen}},\ and\ \bibinfo {author} {\bibfnamefont {S.~N.}\
  \bibnamefont {Yang}},\ }\href {https://doi.org/10.1016/j.physrep.2006.09.006}
  {\bibfield  {journal} {\bibinfo  {journal} {Phys. Rept.}\ }\textbf {\bibinfo
  {volume} {437}},\ \bibinfo {pages} {125} (\bibinfo {year} {2007})},\ \Eprint
  {https://arxiv.org/abs/hep-ph/0609004} {arXiv:hep-ph/0609004 [hep-ph]}
  \BibitemShut {NoStop}%
\bibitem [{\citenamefont {Peskin}\ and\ \citenamefont
  {Schroeder}(1995)}]{pesschr}%
  \BibitemOpen
  \bibfield  {author} {\bibinfo {author} {\bibfnamefont {M.~E.}\ \bibnamefont
  {Peskin}}\ and\ \bibinfo {author} {\bibfnamefont {D.~V.}\ \bibnamefont
  {Schroeder}},\ }\href@noop {} {\emph {\bibinfo {title} {An Introduction to
  Quantum Field Theory}}}\ (\bibinfo  {publisher} {Westview Press},\ \bibinfo
  {year} {1995})\BibitemShut {NoStop}%
\bibitem [{\citenamefont {Weinberg}(1979)}]{Weinberg:1978kz}%
  \BibitemOpen
  \bibfield  {author} {\bibinfo {author} {\bibfnamefont {S.}~\bibnamefont
  {Weinberg}},\ }\href@noop {} {\bibfield  {journal} {\bibinfo  {journal}
  {Physica}\ }\textbf {\bibinfo {volume} {A96}},\ \bibinfo {pages} {327}
  (\bibinfo {year} {1979})}\BibitemShut {NoStop}%
\bibitem [{\citenamefont {Gasser}\ and\ \citenamefont
  {Leutwyler}(1984)}]{Gasser:1983yg}%
  \BibitemOpen
  \bibfield  {author} {\bibinfo {author} {\bibfnamefont {J.}~\bibnamefont
  {Gasser}}\ and\ \bibinfo {author} {\bibfnamefont {H.}~\bibnamefont
  {Leutwyler}},\ }\href {https://doi.org/10.1016/0003-4916(84)90242-2}
  {\bibfield  {journal} {\bibinfo  {journal} {Annals Phys.}\ }\textbf {\bibinfo
  {volume} {158}},\ \bibinfo {pages} {142} (\bibinfo {year}
  {1984})}\BibitemShut {NoStop}%
\bibitem [{\citenamefont {Scherer}(2003)}]{Scherer:2002tk}%
  \BibitemOpen
  \bibfield  {author} {\bibinfo {author} {\bibfnamefont {S.}~\bibnamefont
  {Scherer}},\ }\href@noop {} {\bibfield  {journal} {\bibinfo  {journal} {Adv.
  Nucl. Phys.}\ }\textbf {\bibinfo {volume} {27}},\ \bibinfo {pages} {277}
  (\bibinfo {year} {2003})},\ \Eprint {https://arxiv.org/abs/hep-ph/0210398}
  {arXiv:hep-ph/0210398 [hep-ph]} \BibitemShut {NoStop}%
\bibitem [{\citenamefont {Scherer}\ and\ \citenamefont
  {Schindler}(2012)}]{Scherer:2012xha}%
  \BibitemOpen
  \bibfield  {author} {\bibinfo {author} {\bibfnamefont {S.}~\bibnamefont
  {Scherer}}\ and\ \bibinfo {author} {\bibfnamefont {M.~R.}\ \bibnamefont
  {Schindler}},\ }\bibfield  {journal} {\bibinfo  {journal} {Lect. Notes
  Phys.}\ }\textbf {\bibinfo {volume} {830}},\ \href
  {https://doi.org/10.1007/978-3-642-19254-8} {10.1007/978-3-642-19254-8}
  (\bibinfo {year} {2012})\BibitemShut {NoStop}%
\bibitem [{\citenamefont {Kobayashi}\ and\ \citenamefont
  {Maskawa}(1973)}]{Kobayashi:1973fv}%
  \BibitemOpen
  \bibfield  {author} {\bibinfo {author} {\bibfnamefont {M.}~\bibnamefont
  {Kobayashi}}\ and\ \bibinfo {author} {\bibfnamefont {T.}~\bibnamefont
  {Maskawa}},\ }\href {https://doi.org/10.1143/PTP.49.652} {\bibfield
  {journal} {\bibinfo  {journal} {Prog. Theor. Phys.}\ }\textbf {\bibinfo
  {volume} {49}},\ \bibinfo {pages} {652} (\bibinfo {year} {1973})}\BibitemShut
  {NoStop}%
\bibitem [{\citenamefont {Jenkins}\ and\ \citenamefont
  {Manohar}(1991)}]{Jenkins:1991es}%
  \BibitemOpen
  \bibfield  {author} {\bibinfo {author} {\bibfnamefont {E.~E.}\ \bibnamefont
  {Jenkins}}\ and\ \bibinfo {author} {\bibfnamefont {A.~V.}\ \bibnamefont
  {Manohar}},\ }\href {https://doi.org/10.1016/0370-2693(91)90840-M} {\bibfield
   {journal} {\bibinfo  {journal} {Phys. Lett.}\ }\textbf {\bibinfo {volume}
  {B259}},\ \bibinfo {pages} {353} (\bibinfo {year} {1991})}\BibitemShut
  {NoStop}%
\bibitem [{\citenamefont {Weinberg}(1966)}]{Weinberg:1966kf}%
  \BibitemOpen
  \bibfield  {author} {\bibinfo {author} {\bibfnamefont {S.}~\bibnamefont
  {Weinberg}},\ }\href {https://doi.org/10.1103/PhysRevLett.17.616} {\bibfield
  {journal} {\bibinfo  {journal} {Phys. Rev. Lett.}\ }\textbf {\bibinfo
  {volume} {17}},\ \bibinfo {pages} {616} (\bibinfo {year} {1966})}\BibitemShut
  {NoStop}%
\bibitem [{\citenamefont {Ledwig}\ \emph {et~al.}(2014)\citenamefont {Ledwig},
  \citenamefont {Martin~Camalich}, \citenamefont {Geng},\ and\ \citenamefont
  {Vicente~Vacas}}]{Ledwig:2014rfa}%
  \BibitemOpen
  \bibfield  {author} {\bibinfo {author} {\bibfnamefont {T.}~\bibnamefont
  {Ledwig}}, \bibinfo {author} {\bibfnamefont {J.}~\bibnamefont
  {Martin~Camalich}}, \bibinfo {author} {\bibfnamefont {L.~S.}\ \bibnamefont
  {Geng}},\ and\ \bibinfo {author} {\bibfnamefont {M.~J.}\ \bibnamefont
  {Vicente~Vacas}},\ }\href {https://doi.org/10.1103/PhysRevD.90.054502}
  {\bibfield  {journal} {\bibinfo  {journal} {Phys. Rev.}\ }\textbf {\bibinfo
  {volume} {D90}},\ \bibinfo {pages} {054502} (\bibinfo {year} {2014})},\
  \Eprint {https://arxiv.org/abs/1405.5456} {arXiv:1405.5456 [hep-ph]}
  \BibitemShut {NoStop}%
\bibitem [{\citenamefont {Chivukula}\ and\ \citenamefont
  {Manohar}(1988)}]{Chivukula:1988gp}%
  \BibitemOpen
  \bibfield  {author} {\bibinfo {author} {\bibfnamefont {R.~S.}\ \bibnamefont
  {Chivukula}}\ and\ \bibinfo {author} {\bibfnamefont {A.~V.}\ \bibnamefont
  {Manohar}},\ }\href {https://doi.org/10.1016/0370-2693(88)90891-X} {\bibfield
   {journal} {\bibinfo  {journal} {Phys. Lett. B}\ }\textbf {\bibinfo {volume}
  {207}},\ \bibinfo {pages} {86} (\bibinfo {year} {1988})},\ \bibinfo {note}
  {[Erratum: Phys.Lett.B 217, 568 (1989)]}\BibitemShut {NoStop}%
\bibitem [{\citenamefont {Ledwig}\ \emph {et~al.}(2012)\citenamefont {Ledwig},
  \citenamefont {Martin-Camalich}, \citenamefont {Pascalutsa},\ and\
  \citenamefont {Vanderhaeghen}}]{Ledwig:2011cx}%
  \BibitemOpen
  \bibfield  {author} {\bibinfo {author} {\bibfnamefont {T.}~\bibnamefont
  {Ledwig}}, \bibinfo {author} {\bibfnamefont {J.}~\bibnamefont
  {Martin-Camalich}}, \bibinfo {author} {\bibfnamefont {V.}~\bibnamefont
  {Pascalutsa}},\ and\ \bibinfo {author} {\bibfnamefont {M.}~\bibnamefont
  {Vanderhaeghen}},\ }\href {https://doi.org/10.1103/PhysRevD.85.034013}
  {\bibfield  {journal} {\bibinfo  {journal} {Phys. Rev.}\ }\textbf {\bibinfo
  {volume} {D85}},\ \bibinfo {pages} {034013} (\bibinfo {year} {2012})},\
  \Eprint {https://arxiv.org/abs/1108.2523} {arXiv:1108.2523 [hep-ph]}
  \BibitemShut {NoStop}%
\bibitem [{\citenamefont {Dashen}\ and\ \citenamefont
  {Manohar}(1993)}]{Dashen:1993as}%
  \BibitemOpen
  \bibfield  {author} {\bibinfo {author} {\bibfnamefont {R.~F.}\ \bibnamefont
  {Dashen}}\ and\ \bibinfo {author} {\bibfnamefont {A.~V.}\ \bibnamefont
  {Manohar}},\ }\href {https://doi.org/10.1016/0370-2693(93)91635-Z} {\bibfield
   {journal} {\bibinfo  {journal} {Phys. Lett.}\ }\textbf {\bibinfo {volume}
  {B315}},\ \bibinfo {pages} {425} (\bibinfo {year} {1993})},\ \Eprint
  {https://arxiv.org/abs/hep-ph/9307241} {arXiv:hep-ph/9307241 [hep-ph]}
  \BibitemShut {NoStop}%
\bibitem [{\citenamefont {{Wolfram Research{,} Inc.}}(2021)}]{Mathematica}%
  \BibitemOpen
  \bibfield  {author} {\bibinfo {author} {\bibnamefont {{Wolfram Research{,}
  Inc.}}},\ }\href {https://www.wolfram.com/mathematica} {\bibinfo {title}
  {Mathematica, {V}ersion 12.3}} (\bibinfo {year} {2021}),\ \bibinfo {note}
  {champaign, IL, USA}\BibitemShut {NoStop}%
\bibitem [{\citenamefont {Mertig}\ \emph {et~al.}(1991)\citenamefont {Mertig},
  \citenamefont {Bohm},\ and\ \citenamefont {Denner}}]{Mertig:1990an}%
  \BibitemOpen
  \bibfield  {author} {\bibinfo {author} {\bibfnamefont {R.}~\bibnamefont
  {Mertig}}, \bibinfo {author} {\bibfnamefont {M.}~\bibnamefont {Bohm}},\ and\
  \bibinfo {author} {\bibfnamefont {A.}~\bibnamefont {Denner}},\ }\href
  {https://doi.org/10.1016/0010-4655(91)90130-D} {\bibfield  {journal}
  {\bibinfo  {journal} {Comput. Phys. Commun.}\ }\textbf {\bibinfo {volume}
  {64}},\ \bibinfo {pages} {345} (\bibinfo {year} {1991})}\BibitemShut
  {NoStop}%
\bibitem [{\citenamefont {Shtabovenko}\ \emph {et~al.}(2016)\citenamefont
  {Shtabovenko}, \citenamefont {Mertig},\ and\ \citenamefont
  {Orellana}}]{Shtabovenko:2016sxi}%
  \BibitemOpen
  \bibfield  {author} {\bibinfo {author} {\bibfnamefont {V.}~\bibnamefont
  {Shtabovenko}}, \bibinfo {author} {\bibfnamefont {R.}~\bibnamefont
  {Mertig}},\ and\ \bibinfo {author} {\bibfnamefont {F.}~\bibnamefont
  {Orellana}},\ }\href {https://doi.org/10.1016/j.cpc.2016.06.008} {\bibfield
  {journal} {\bibinfo  {journal} {Comput. Phys. Commun.}\ }\textbf {\bibinfo
  {volume} {207}},\ \bibinfo {pages} {432} (\bibinfo {year} {2016})},\ \Eprint
  {https://arxiv.org/abs/1601.01167} {arXiv:1601.01167 [hep-ph]} \BibitemShut
  {NoStop}%
\bibitem [{\citenamefont {Shtabovenko}\ \emph {et~al.}(2020)\citenamefont
  {Shtabovenko}, \citenamefont {Mertig},\ and\ \citenamefont
  {Orellana}}]{Shtabovenko:2020gxv}%
  \BibitemOpen
  \bibfield  {author} {\bibinfo {author} {\bibfnamefont {V.}~\bibnamefont
  {Shtabovenko}}, \bibinfo {author} {\bibfnamefont {R.}~\bibnamefont
  {Mertig}},\ and\ \bibinfo {author} {\bibfnamefont {F.}~\bibnamefont
  {Orellana}},\ }\href {https://doi.org/10.1016/j.cpc.2020.107478} {\bibfield
  {journal} {\bibinfo  {journal} {Comput. Phys. Commun.}\ }\textbf {\bibinfo
  {volume} {256}},\ \bibinfo {pages} {107478} (\bibinfo {year} {2020})},\
  \Eprint {https://arxiv.org/abs/2001.04407} {arXiv:2001.04407 [hep-ph]}
  \BibitemShut {NoStop}%
\bibitem [{\citenamefont {Mommers}(2022)}]{Mommers1677385}%
  \BibitemOpen
  \bibfield  {author} {\bibinfo {author} {\bibfnamefont {C.~J.}\ \bibnamefont
  {Mommers}},\ }\emph {\bibinfo {title} {Rare Weak Decays of the Omega
  Baryon}},\ \href@noop {} {Master's thesis},\ \bibinfo  {school} {Uppsala
  University, Nuclear Physics} (\bibinfo {year} {2022})\BibitemShut {NoStop}%
\bibitem [{\citenamefont {Borasoy}\ and\ \citenamefont
  {Holstein}(1999)}]{Borasoy:1998ku}%
  \BibitemOpen
  \bibfield  {author} {\bibinfo {author} {\bibfnamefont {B.}~\bibnamefont
  {Borasoy}}\ and\ \bibinfo {author} {\bibfnamefont {B.~R.}\ \bibnamefont
  {Holstein}},\ }\href {https://doi.org/10.1007/s100520050323} {\bibfield
  {journal} {\bibinfo  {journal} {Eur. Phys. J. C}\ }\textbf {\bibinfo {volume}
  {6}},\ \bibinfo {pages} {85} (\bibinfo {year} {1999})},\ \Eprint
  {https://arxiv.org/abs/hep-ph/9805430} {arXiv:hep-ph/9805430} \BibitemShut
  {NoStop}%
\bibitem [{\citenamefont {Hemmert}\ \emph {et~al.}(1998)\citenamefont
  {Hemmert}, \citenamefont {Holstein},\ and\ \citenamefont
  {Kambor}}]{Hemmert:1997ye}%
  \BibitemOpen
  \bibfield  {author} {\bibinfo {author} {\bibfnamefont {T.~R.}\ \bibnamefont
  {Hemmert}}, \bibinfo {author} {\bibfnamefont {B.~R.}\ \bibnamefont
  {Holstein}},\ and\ \bibinfo {author} {\bibfnamefont {J.}~\bibnamefont
  {Kambor}},\ }\href {https://doi.org/10.1088/0954-3899/24/10/003} {\bibfield
  {journal} {\bibinfo  {journal} {J. Phys. G}\ }\textbf {\bibinfo {volume}
  {24}},\ \bibinfo {pages} {1831} (\bibinfo {year} {1998})},\ \Eprint
  {https://arxiv.org/abs/hep-ph/9712496} {arXiv:hep-ph/9712496} \BibitemShut
  {NoStop}%
\bibitem [{\citenamefont {Rarita}\ and\ \citenamefont
  {Schwinger}(1941)}]{Rarita:1941mf}%
  \BibitemOpen
  \bibfield  {author} {\bibinfo {author} {\bibfnamefont {W.}~\bibnamefont
  {Rarita}}\ and\ \bibinfo {author} {\bibfnamefont {J.}~\bibnamefont
  {Schwinger}},\ }\href {https://doi.org/10.1103/PhysRev.60.61} {\bibfield
  {journal} {\bibinfo  {journal} {Phys. Rev.}\ }\textbf {\bibinfo {volume}
  {60}},\ \bibinfo {pages} {61} (\bibinfo {year} {1941})}\BibitemShut {NoStop}%
\bibitem [{\citenamefont {de~Jong}\ and\ \citenamefont
  {Malfliet}(1992)}]{deJong:1992wm}%
  \BibitemOpen
  \bibfield  {author} {\bibinfo {author} {\bibfnamefont {F.}~\bibnamefont
  {de~Jong}}\ and\ \bibinfo {author} {\bibfnamefont {R.}~\bibnamefont
  {Malfliet}},\ }\href {https://doi.org/10.1103/PhysRevC.46.2567} {\bibfield
  {journal} {\bibinfo  {journal} {Phys. Rev.}\ }\textbf {\bibinfo {volume}
  {C46}},\ \bibinfo {pages} {2567} (\bibinfo {year} {1992})}\BibitemShut
  {NoStop}%
\bibitem [{\citenamefont {Holmberg}(2019)}]{Holmberg1372811}%
  \BibitemOpen
  \bibfield  {author} {\bibinfo {author} {\bibfnamefont {M.}~\bibnamefont
  {Holmberg}},\ }\emph {\bibinfo {title} {Detemining Low-Energy Constants in
  $\chi$PT From Decays of Decuplet Baryons}},\ \href@noop {} {Master's
  thesis},\ \bibinfo  {school} {Uppsala University, Nuclear Physics} (\bibinfo
  {year} {2019})\BibitemShut {NoStop}%
\end{thebibliography}%

\end{document}